\documentclass[twocolumn]{aastex62}
\usepackage{graphicx}
\usepackage{subfigure}
\usepackage{epstopdf}
\usepackage{amsmath} 
\usepackage{amsmath}
\usepackage{bm}
\usepackage{graphicx}
\usepackage{multirow}
\usepackage{enumerate}
\usepackage{amsmath}
\usepackage{bm}
\usepackage{graphicx}
\usepackage{natbib}
\usepackage{color}
\hypersetup{colorlinks,
linkcolor=blue,
citecolor=blue,
filecolor=blue,
urlcolor=blue}

\usepackage{soul}

\usepackage{enumerate}
\usepackage{textcomp}

\usepackage{multirow}
\usepackage{enumerate}

\begin{document}
\title{Evolution of spiral galaxies in nonlocal gravity}
\author{Mahmood Roshan}
\affiliation{Department of Physics, Ferdowsi University of Mashhad, P.O.Box 1436, Mashhad, Iran}
\email{mroshan@um.ac.ir}
\author{Sohrab Rahvar}
\affiliation{Physics Department, Sharif University, P.O.Box 11365-9161, Azadi Avenue, Tehran, Iran}
\email{rahvar@sharif.edu}
\begin{abstract}
We study the evolution of simulated disk galaxies in the context of a nonlocal theory of gravity. In this theory the appearance of the dark matter problem in cosmology and astrophysics is a manifestation of the nonlocality of the gravitational interaction. Using high-resolution N-body simulations, we investigate the dynamical evolution of disk galaxies and compare the result with the standard dark matter viewpoint. Specifically, we construct two exponential galaxy models, one in nonlocal gravity (NLG) and the other surrounded by a Plummer dark matter halo. Both systems start from the same baryonic matter distribution, particles velocities, random velocities and the initial Toomre's parameter. However, although the initial conditions are the same in both models, their long-term dynamics reveals some notable differences. For example, it turns out that the bar instability happens with higher rate in NLG model compared with the standard case. On the other hand, at the end of the simulation, we find that bars are weaker and faster in NLG compared with the standard case.
\end{abstract}

\keywords{galaxies: kinematics and dynamics-- galaxies: spiral-- instabilities-- galaxies: bar growth}

\section{\small{Introduction}}
\label{introduction}
Early numerical simulations revealed that disk galaxies are violently unstable against stellar bar formation, for example, see \citet{miller}; \citet{ho}. This rapid bar formation was inconsistent with the observations in the 1960s. However, it turned out that a rigid spherical halo surrounding the galaxy, can substantially suppress the bar instability \citep{op}. Currently, it is well-known that disks in live dark matter halos are more prone to bar instability compared to those in rigid halos \citep{at2002,se16}.

The dark matter particles distributed in a live halo can effectively slow down the bar pattern speed. More specifically the \textit{dynamical friction} originated from gravitational interaction between the dark matter particles and the baryonic matter in the disk, is not negligible. This mechanism leads to slow bars at the end of dark matter halo simulations, for example in the cosmological context see \cite{cosmo} and for isolated galaxies see \cite{deb2000}.  However, this fact is inconsistent with the relevant observations on the pattern speeds. Theses observations show that spiral galaxies host fast bars \citep{pattern}. It should be noted that, for a fast bar the ratio of the corotation radius to the bar semi-major axis is smaller than $1.4$. This important issue will be discussed at greater length in the subsequent sections.

These problems along with other similar unresolved issues in the standard dark matter paradigm give rise to motivations for modified gravity approach. In this paper we study the galaxy evolution in a nonlocal gravity theory \citep{mashhoon,mashhoonbook} which tries to address the dark matter problem without invoking exotic particles \citep{rahvar}. Therefore, let us briefly review the similar studies in the context of other well-known modified gravity theories. 

 Modified Newtonian Dynamics (MOND) is one of the most successful theories to address the dark matter problem (\citealt{milgrom}; \citealt{fa}). The local stability of disk galaxies in MOND has been studied in \cite{mil89}. There are few papers in which the evolution of spiral galaxies, with emphasis on the bar instability, has been investigated via N-body simulations, for example see \citet{chris}; \citet{brada} and \citet{ti}. The Latter provides more advanced simulations incorporating a large number of particles in the simulations. Therefore its results are more reliable than precedent papers. In this paper,  it has been shown that the bar growth rate in MOND is higher than in Newtonian case. However, in MOND after reaching a maximum, the bar strength starts to decrease while in the dark matter model it does not stop increasing. In this case, the final magnitude of the bar at the end of the simulation in MOND is substantially smaller than the dark matter halo case. 

Another theory which explains the flat rotation curves and the mass discrepancy in the clusters is a scalar-tensor-vector theory of gravity known as MOG in the literature \citep{m2006,2013MNRAS.436.1439M,2014MNRAS.441.3724M}. The local stability of disk galaxies in MOG has been investigated in \cite{ro2015}. On the other hand, the global stability of spiral galaxies has been studied in \cite{gr2017} and \cite{roshan2018}. Both studies, using different codes, show that the bar growth rate is lower in MOG. Furthermore, the magnitude of the bar at the end of the simulation is smaller in MOG, and the pattern speed is faster than that of the standard halo model. The absence of the dynamical friction in MOG is the main reason for the latter case.

In this paper, we study the time evolution of exponential disk galaxies in the context of the nonlocal theory of gravity (NLG). This theory has been built on completely different bases compared with MOND and MOG but shares the same purpose to resolve the dark matter problem without introducing dark matter particles \citep{mashhoon,mashhoonbook}. In this theory, gravity is similar to the electrodynamics in the non-vacuum medium and has a non-local behavior. We know this behavior from the concept of hysteresis in the magnetic materials \citep{jackson}.

The non-locality in this modified gravity model heritages from the violation of the equivalence principle where it postulates a pointwise 
connection between the local frame of the accelerated observer with the background global inertial frame. Writing the general relativity in non-local form, in the weak field regime, we can write the modified Poisson equation where the potential from the all the space with a given kernel is engaged in this equation. The comparison of the dynamics from the NLG with the observation of galaxies and cluster of galaxies can explain the dynamics of large-scale structures without a need to the dark matter \citep{rahvar}.

The outline of this paper is as follows: In section \ref{wfl} we briefly introduce NLG and discuss its weak filed limit. In section \ref{gi} we explain the numerical procedure to produce the initial conditions. Moreover we describe the essential ingredients of the code which evolves the point particles. We construct two models with identical initial conditions, one model in NLG without halo component, and the other in Newtonian gravity surrounded by a live dark matter halo. In section \ref{res} we compare the galactic evolution in these models. Finally, results are summarized and discussed in section \ref{disc}.

\section{\small{Nonlocal gravity: the weak field limit}}
\label{wfl}
It is natural to expect that relativistic effects are negligible in the secular dynamics of spiral galaxies. Excluding the central black hole, this fact is true almost everywhere in the galaxy. Therefore we need to use the weak-field limit of the theory. More specifically, in N-body simulations, we need the gravitational force between two point particles. In this section, we review the main features of NLG and its Newtonian limit.

The Lorentz invariance between the inertial observers is the central symmetry of the special relativity. In this theory, all the measurements are done locally both in the space and in the time. The idea of non-locality dates back to Bohr and Rosenfeld where they pointed out long ago that field determination might not be performed instantaneously \citep{bohr}. This is a covariant theory, however the average value of the electromagnetic field is detemined in a finite volume of the space-time as 
\begin{equation}
F_{\mu\nu}(R) = \frac{1}{R} \int_R F_{\mu\nu}(x')dx'^4,
\end{equation}
where the size of space-time volume of $R$ might be defined by the natural volume of space as $(\hbar/mc)^3$ and time $\hbar/mc^2$. 

There are also classical examples of non-locality in the electrodynamics \citep{jackson}. For the case that the dielectric constant in a medium depends on the frequency of an external electric field, the displacement vector relates to the electric field as 
\begin{equation}
D(\bf {x},\omega) = \epsilon(\omega)E({\bf x},\omega).
\label{dd}
\end{equation}
The Fourier transformation of this equation results in 
\begin{equation}
D({\bf x},t) = E({\bf x},t) + \int_{-\infty}^{+\infty} G(\tau)E({\bf x},t-\tau) d\tau,
\end{equation}
where $G(\tau) = 1/2\pi \int^\infty_{-\infty} \left(\epsilon(\omega)-1\right) e^{i\omega\tau} d\omega$ and depending on the dielectric function, one can find the non-locality of the displacement vector in time (for detail see Chapter 7 of \citealt{jackson}). The extension of equation (\ref{dd}) is 
\begin{equation}
D^i({\bf k},\omega) = \sum_j \epsilon^{ij}({\bf k},\omega) E^i({\bf k},\omega)
\end{equation}
and in the Fourier space results in non-locality both in the space and time as
\begin{equation}
D^i({\bf x},t) = \sum_j \int dx'^3\int dt' \epsilon^{ij}({\bf x'},t') E^j({\bf x-x'},t-t'). 
\end{equation}
In the superconductors where the mean-free path is comparable with the size of system, the Ohm's law is 
no longer valid and dielectric tensor not only depends on the frequency but also it depends on 
the wavenumber.

The idea of non-locality is extended to the gravitation where for the non-inertial observers unlike to the general relativity, the equivalence of a local accelerating observer to an inertial observer is not valid. Here, we define a characteristic length of the system by $\lambda$ and a characteristic length for the acceleration of a reference frame by $L$ where the ratio of $\lambda/L$ would provide the degree of the non-locality. 

 We can define the characteristic length of $L$  by $L=c^2/a$ for linear accelerating frames and $L=c/\Omega$ for rotating systems. Now, let us calculate  $\lambda/L$ for typical astrophysical and cosmological systems. (i) For the case of Earth using $a=g$ results in $\lambda/L \simeq 10^{-12}$. (ii) For the case of galaxy with the rotating velocity of $v_{rot}$, $L = c\lambda/v_{rot}$ where $\lambda$ is the size of galaxy, then $\lambda/L =v_{rot}/c \simeq 10^{-3}$. (iii), finally for the case of the Universe, the non-locality ratio is $\lambda/L = \frac12 (HR/c)^2$ where $H$ is the Hubble parameter and $R$ is scale of Universe in our concern. From these three scales we can see that for the larger scales and/or strong gravitational system, the non-locality is important as this effect depends on the combination of scale and the acceleration. The NLG can be written in terms of the teleparallel gravity theories since the final formalism is like the electromagnetism and we know how to make the electromagnetism non-local in a medium \citep{jackson}. 

The Einstein equations can be written in equivalent formalism so-called teleparallel gravity, namely, $GR_{\parallel}$. In this formalism the Riemannian structure of space-time has an additional Weitzenb{\"o}ck connection. The advantage of this formalism is that Einstein gravity 
equations are expressed in the form of covariant Maxwell equations. In other word, in this formalism general relativity is a gauge theory of the Abelian group of spacetime translations \citep{tpar}. The advantage of this formalism is that in analogy to the electromagnetism, we  can make these theories non-local. The gravitational potential in teleparallel gravity is given by the tetrad field which relates the metric of space-time to the local tangent Minkowski space  as $g_{\mu\nu}(x) = e_{\mu}{}^{\hat\alpha}(x)e_{\nu}{}^{\hat\beta}(x) \eta_{\alpha\beta}$ where the connection is defined by $\Gamma^{\beta}{}_{\mu\nu} = e^{\beta}{}_\alpha\partial_\nu e^\alpha{}_\mu$. In this definition the tetrad field is globally teleparallel and the curvature of the 
 Weitzenb{\"o}ck  spacetime vanishes. The field equations in this formalism are expressed in terms of the gravitational field strength $C_{\mu\nu}{}^{\hat\alpha} = \partial_\mu e_\nu{}^{\hat\alpha} - \partial_\nu e_\mu{}^{\hat\alpha}$ which is similar to the electromagnetic tensor of $F_{\mu\nu} = \partial_\mu A_\nu - \partial_\nu A_\mu$. The result is similar to the Maxwell equations where according to the non-localized electromagnetic field, we can make gravity non-local in this formalism (for details see \citealt{mashhoonbook}). Actually we 
 don't have a microphysical description of the non-locality in GR unlike to what we have in the electromagnetism. However,  we may assume the non-locality of GR as an emergent phenomenon from the quantum behavior of gravity, similar to what we have from the microphysics of the dielectric medium.  
 
 For the weak field approximation the tetrad field can be expanded around the flat space as $e^{\hat\alpha}{}_\mu = \delta^{\alpha}{}_{\mu} + \psi^\alpha{}_{\mu}$ where since $ \psi^\alpha{}_{\mu}$ is the first order perturbation and in this approximation, there is no distinction between the anholonomic and holonomic indices. After detailed calculation, the modified Poisson equation in this theory can be written as
 \begin{equation}
\nabla^2\phi({\bf x}) = 4\pi G\rho({\bf x}) + 4\pi G \int q(\bf{x-y})\rho({\bf y}) d^3y,
\label{phi} 
 \end{equation} 
where for $q(\bf{x-y}) = 0$ we have locality and recover the standard Poisson equation. The important characteristics of this equation is that for NLG the second term at the right hand side plays the role of dark matter. Moreover, the linearity of this equation in terms of the source is satisfied. This means that we can use the superposition of discrete sources for calculation of the overall potential for an N-body system. The kernel function has no fundamental definition and can be obtained from comparing the observational data and theory. 

Here we use the kernel function of 
\begin{equation}
q(r) = \frac{1}{4\pi\lambda_0}\frac{1+\mu r}{r^2}e^{-\mu r},
\end{equation}
where $\lambda_0$ and $\mu$ are the constant parameters of this function. We substitute this kernel function in equation (\ref{phi}) and use Dirac-Delta function as a point-like source of gravity, the result from this integration is
\begin{equation}
\phi(r)=-\frac{G m}{r}\Big\lbrace1+\alpha(1-e^{-\mu r})-\frac{\mu\alpha r}{2}\int_{\infty}^{r}\frac{e^{-\mu r'}}{r'}dr'\Big\rbrace,
\label{potnlg}
\end{equation}
where we replace $\lambda_0$ with $\lambda_0 = 2/(\alpha\mu)$ and taking derivative from this equation we obtain the acceleration of a test mass particle as follows: 
\begin{equation}\label{force}
a(r)=\frac{Gm}{r^2}\Big\lbrace1+\alpha\Big(1-\Big(1+\frac{\mu r}{2}\Big)e^{-\mu r}\Big).\Big\rbrace
\end{equation}
We note that due the linearity of Poisson equation in terms of density, we can use the superposition of the individual particles on the overall gravitational acceleration. 

This theory has been used for calculation of the acceleration of a test particle in the spiral galaxies \citep{rahvar}. The results from this analysis show the unique value of $\alpha = 10.94\pm 2.56$ and $\mu = 0.059\pm 0.028~{\text kpc}^{-1}$ can be used to describe the dynamics of these galaxies without need to the dark matter.  Moreover, taking the adapted values for the parameters of this modified gravity model, we can also explain the dynamics of the cluster of galaxies without a need to the contribution of the dark matter.

\section{\small{Numerical method}}
\label{gi}
One of the standard and high-resolution N-body codes which has been developed to study isolated galaxies is the \textit{GALAXY code} \cite{se2014}. This code produces suitable initial conditions and gives the time evolution of the galactic models. The galactic models can include disk, bulge, and spherical dark matter halo. Almost all the well-known disk and halo density profiles are included in the code. A detailed description is available in the online manual \cite{se2014}. We recall that in NLG model, we do not need dark matter halos. Therefore to construct galactic models in this theory, we have modified only some parts of the code which deal with disks. As a necessary test, we measure the error in total energy during the simulations. This error is always smaller than $1$\% in all the models. For simplicity, we have not included a bulge in our models.

For galaxy model in Newtonian gravity including a live halo, we use the hybrid N-body code described in \citep[Appendix B]{se3}. In this case two grids are used to find the gravitational field: a spherical grid for the live halo component, and a 3D cylindrical polar grid for the disk. Of course, our single component model in NLG uses only the latter grid.

\subsection{Initial conditions}

The initial dark matter model in Newtonian gravity is identical to the model "EPL"\footnote{E: Exponential disk, P: Plummer sphere, L: Live halo} in \cite{roshan2018}. This model is consist of an exponential disk and a live Plummer spherical halo.  The exponential profile for the disk is a satisfactory choice from the observational point of view. On the other hand, the Plummer sphere for the halo is chosen to obtain a better match between rotation curves of this model and the model in NLG. Moreover, this halo has been used in other studies where MOND's consequences are compared with the standard dark matter paradigm, for example see \cite{ti} and \cite{com1}. Therefore, this profile enables us to compare our simulations in NLG with those of MOND. On the other hand, this halo provides rising rotation curves reminiscent of the late-type low luminosity galaxies. The exponential disk and the Plummer halo are given by the following densities respectively
\begin{eqnarray}
\Sigma(R)=\frac{M_{d}}{2\pi R_d^2}\,e^{-R/R_d}
\end{eqnarray}
\begin{eqnarray}
\rho(r)=\frac{3 M_h}{4\pi b^3}\Big[1+\Big(\frac{r}{b}\Big)^2\Big]^{-5/2}
\end{eqnarray}
where $M_d$ is the disk mass, and $M_h$ is the halo mass. Furthermore, $R_d$ is disk length scale, and $b$ is a characteristic radius. The disk is truncated at $r=4\, R_d$ and we use a cubic function in order to taper the exponential density smoothly to zero over the range $3.2\, R_d < R < 4\, R_d$. The halo is truncated at $r= 3\, b$. We set the mass of the disk and the halo as $M_d$, and $8 \, M_d$ respectively. Furthermore we assume that $b=10\,R_d$. Notice that we use units such that $R_d=M_d=1$. The Newtonian gravitational constant is scaled as $G=1$. With these choices, the unit of velocity is $V_0=(G M_d/R_d)^{1/2}$, and the time unit is written as $\tau_0=R_d/V_0=(R_d^3/G M)^{1/2}$. In this unit, a suitable choice for our models can be $R_d=2.6$ kpc and $\tau_0=10$ Myr, which yields $M_d\simeq 4 \times 10^{10} M_{\odot}$ and $V_0\simeq 254\, \text{km}/\text{s}$. 

In this paper, the time duration of the simulations is $800 \tau_0$, and in few cases $500 \tau_0$. The time step is $\Delta t= 0.01 \tau_0$. To obtain a better efficiency the time step is increased by successive factors of $2$ in three radial zones. Moreover to make sure that there is no dependence in the outcome on the time step $\Delta t$, we use both shorter and longer time steps, to demonstrate that our choice is appropriate. In all the models we use the cubic spline softening length \citep{mon1992} $\epsilon=0.16\,R_d$. To avoid numeric artifacts the grid is recentered every $16$ time step.

Since the initial matter distribution has been determined, the rotation curve, $v_c$, can be simply calculated. Let us briefly discuss the corresponding velocity dispersions. It is natural to expect that the particles, at least at $\tau=0$, move on nearly circular orbits. Although in real spiral galaxies, most of the stars are supposed to travel on nearly circular orbits, we are not constructing initial conditions identical to a real galaxy. Rather we construct a suitable background disk for studying its global stability in a transparent way. In other words, the simulations start with a disk of particles supported in equilibrium almost entirely by rotation. This cold disk enables us to observe the occurrence and evolution of the stellar bar as a pressure supported configuration. Consequently, there is a relation between radial and azimuthal dispersions as $\sigma_{\phi}=\kappa \sigma_R/2\Omega$, where $\Omega(R)=v_c/R$. On the other hand, working in the epicyclic approximation, we set the radial velocity dispersion $\sigma_R$ in such a way to establish the local stability criterion $Q>1$.  Where $Q$ is the so-called Toomre's stability parameter defined as $Q=\sigma_R/\sigma_{R,\text{crit}}$, where $\sigma_{R,\text{crit}}=3.36 G \Sigma/\kappa$ and $\kappa$ is the epicyclic frequency \citep{toom64}. We set $Q\simeq1.5$ everywhere on the surface of the disk. This fixes the radial dependence of both radial and azimuthal velocity dispersions.

It is necessary to mention that the disk has a Gaussian density profile in the vertical direction $z$ with the scale height $z_0=0.05 R_d$. To find the vertical velocities, the vertical one dimensional Jeans equation is integrated as
\begin{eqnarray}
\sigma_z^2(R,z)=\frac{1}{\rho(R,z)}\int_z^{\infty}\rho(R,z')\frac{\partial \Phi}{\partial z'}dz'.
\end{eqnarray}

Notice that the isotropic distribution function (DF), see \cite{de} for more details, has been used for the Plummer halo in our simulations. However, when the halo is combined with the exponential disk, one needs to recompute the equilibrium DF for the composite system. The \textit{GALAXY} code  uses a halo compression algorithm described by \cite{y1980} and \cite{sm2005}. The algorithm is based on the fact that both radial action and the total angular momentum are conserved during compression. Finally, the particles are selected from the new DF function following the procedure studied in \cite{deb2000}.

Now, the initial conditions for the numeric galactic model in Newtonian gravity have been completely determined. Let us briefly discuss the model in NLG. Hereafter we call this model as "ENLG" (Exponential disk in NLG). As already mentioned, there is no dark matter halo in this model. Therefore, since we have restricted ourselves to the bulge-less simulations, ENLG is a single component disk model. The disk is exponential and its physical parameters are the same as those of the model EPL. The only difference is that the gravitational force between particles is different from the standard case and is given by \eqref{force}. More specifically, since the theory has two free parameters $\alpha$ and $\mu$, we are able to reproduce almost the same initial conditions produced with the two-component model EPL. However, as we shall see, the long-term evolution of the models faces significant differences. 
 
As our final remark in this section, notice that for each responsive component we employ three million particles. For example, in EPL we employed three million particles for the disk and accordingly three million particles for the live halo. Similarly, the ENLG model has three million particles. In order to ensure that the results do not depend on the number of particles, we increase the number of particles up to $N=2\times 10^7$.
\begin{figure} 
 \centerline{\includegraphics[width=8.4cm]{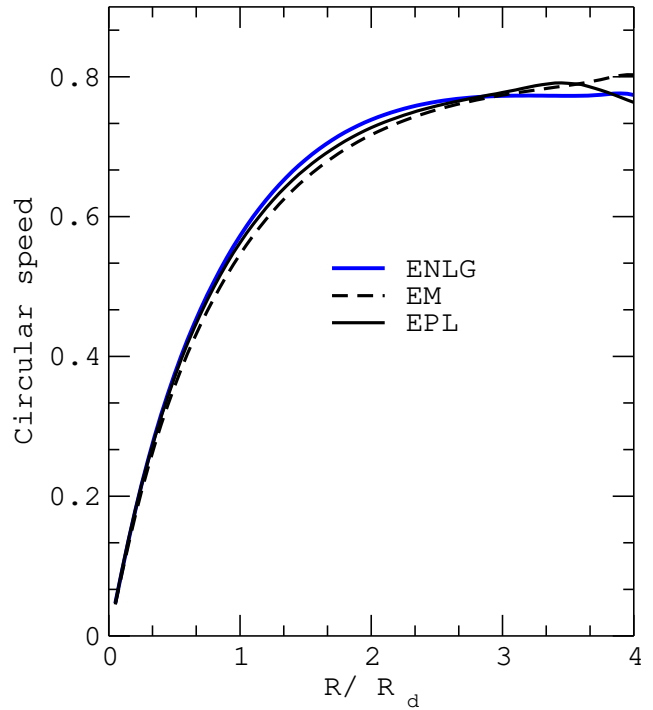}}
\caption{Initial rotational velocities for our models. The blue curve belongs to the model ENLG in NLG. The solid black curve is the rotation curve in the EPL model. Furthermore, the dashed curve belongs to the model EM for MOG studied in \cite{roshan2018}. The model EM is presented just for comparison. For the model ENLG, we have set $\alpha=10.9$ and $\mu=4.5\times 10^{-3} R_{d}^{-1}$. }
\label{vc}
\end{figure}

\subsection{Rotation curves}
Working in the units introduced in the previous subsection, we have plotted the rotation curves in  Fig. \ref{vc}. The blue curve belongs to the model ENLG, and the solid black curve is the corresponding curve for the model EPL. Furthermore, just for comparison, we have shown the rotation curve of model EM, introduced in \cite{roshan2018}, with a dashed curve. This model is a single component exponential disk in MOG. The gravitational force in MOG and NLG includes similar correction terms. Therefore one may expect that these theories predict similar evolution for disk galaxies. However, as we shall show, this is not the case. 

We emphasize again that for a meaningful comparison between different theories, the galactic models should start with the same initial conditions. From Fig. \ref{vc}, we see that the initial rotation curves are similar. Although we have not plotted the initial velocity dispersions, and other functions like $Q(R)$, they also follow similar profiles. 

For a given $\alpha$ and $\mu$ one may first find the rotation curve of the ENLG model, and then change the halo parameters in EPL model until the corresponding rotation curve gets almost identical to that of ENLG model. Of course, one can first find the rotation curve of EPL, and then vary $\alpha$ and $\mu$ in the model ENLG to fit the rotation curve. Using the second approach, we have fixed $\alpha=10.9$ and $\mu=4.5\times 10^{-3} R_{d}^{-1}$ in our simulations. Notice that although the $\alpha$ parameter is close to the observed values, we do not have to use observational values. In fact we have fixed $\alpha$ and $\mu$ by fitting the rotation curve of a numeric model EPL. Consequently, we do not expect that these values get identical to those obtained from fittings to real rotation curves.
\begin{figure}
 \center
\centerline{\includegraphics[width=8.5cm]{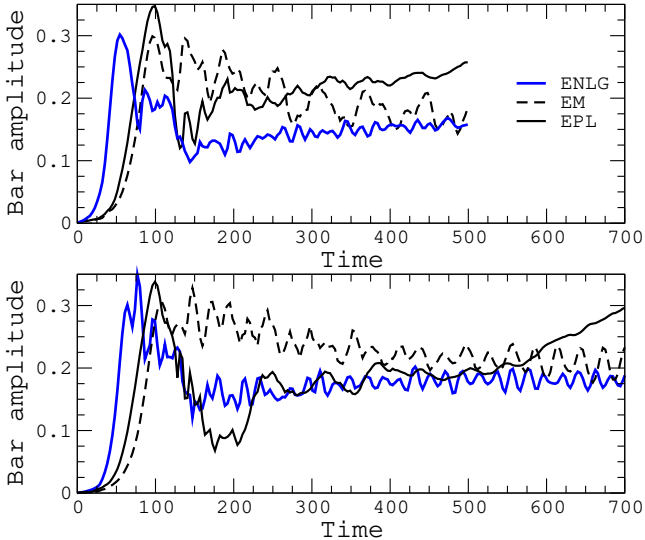}}
\caption{The top panel shows the evolution of the bar amplitude for models EPL, ENLG and EM. In this panel $N=3\times 10^6$. To ensure that the results do not depend on the number of particles, in the bottom panel the bar magnitude is computed for $N=2\times 10^7$ in each responsive component.}
\label{bar}
\end{figure}

\section{\small{Results}}
\label{res}
With the full set of initial conditions in hand, we are ready to discuss the time evolution of the models. Notice that the \textit{GALAXY} code determines the gravitational forces on the cylindrical polar grid using sectoral harmonics $0\leq m\leq 8$. On the other hand, it uses the surface harmonics $0\leq l\leq 4$ to compute the gravitational force on the spherical grid. Albeit it should be noted that this numbers can be changed with the user. We emphasize that for our main model in this paper, i.e., the model ENLG, we have used the softening length $\epsilon=0.16\,R_d$ and grid of $193\times224\times 125$ mesh points. In all the models the grid has been regulated to be large enough to prevent particles from escaping the grid. Moreover to make sure that the results are insensitive to our choice for $\Delta t$, we used shorter and longer values for the time step as $\Delta t=0.005$, $0.01$ and $0.03$. The results are qualitatively the same. Therefore we report the results obtained with $\Delta t=0.01$.

As already mentioned there are two main quantities which are of special importance for us, the stellar bar amplitude and the pattern speed of the bar. Both quantities are sensitive to the existence of the dark matter halo. In other words their time evolution is correlated with the physical properties of the halo. Therefore it is natural to expect serious differences in theories where there is no dark matter halo. In the following subsections, we study these quantities with details in the context of NLG and compare them with those in MOG and the standard dark matter model.
\subsection{Bar growth}

 \begin{figure*}
\centerline{\includegraphics[width=5.5cm]{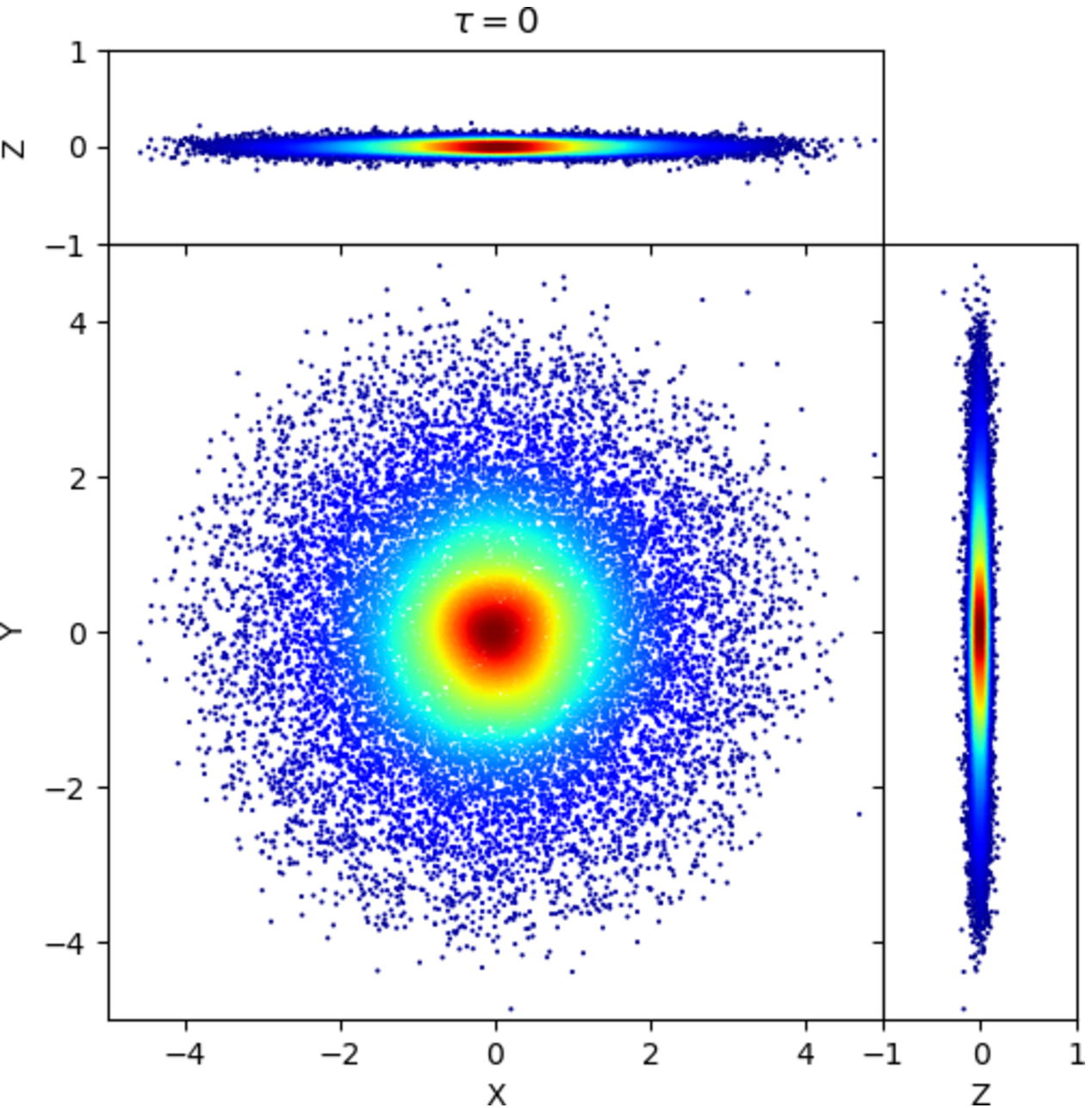}\hspace{0.01 cm} \includegraphics[width=5.5cm]{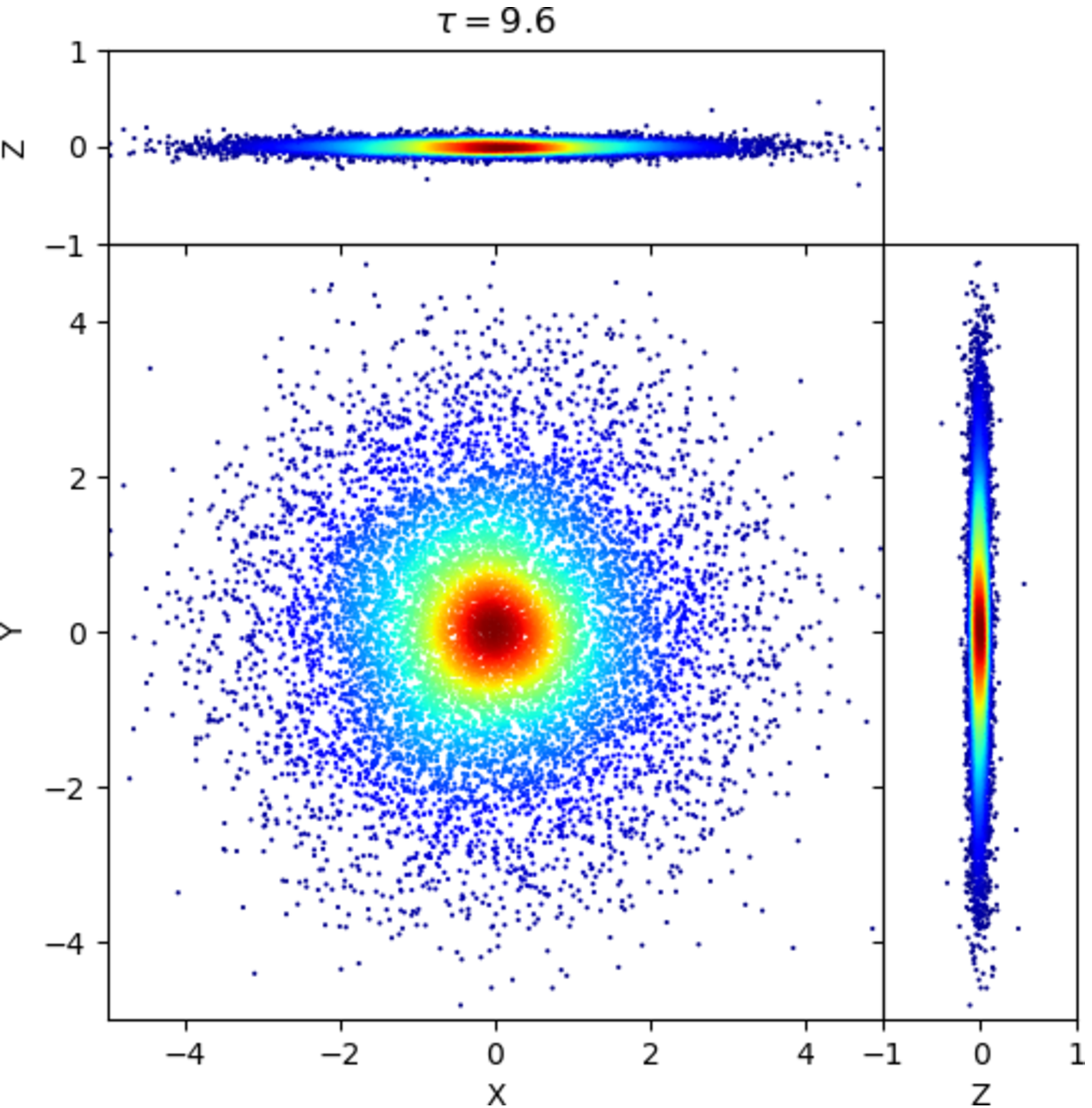}\hspace{0.01 cm} \includegraphics[width=5.5cm]{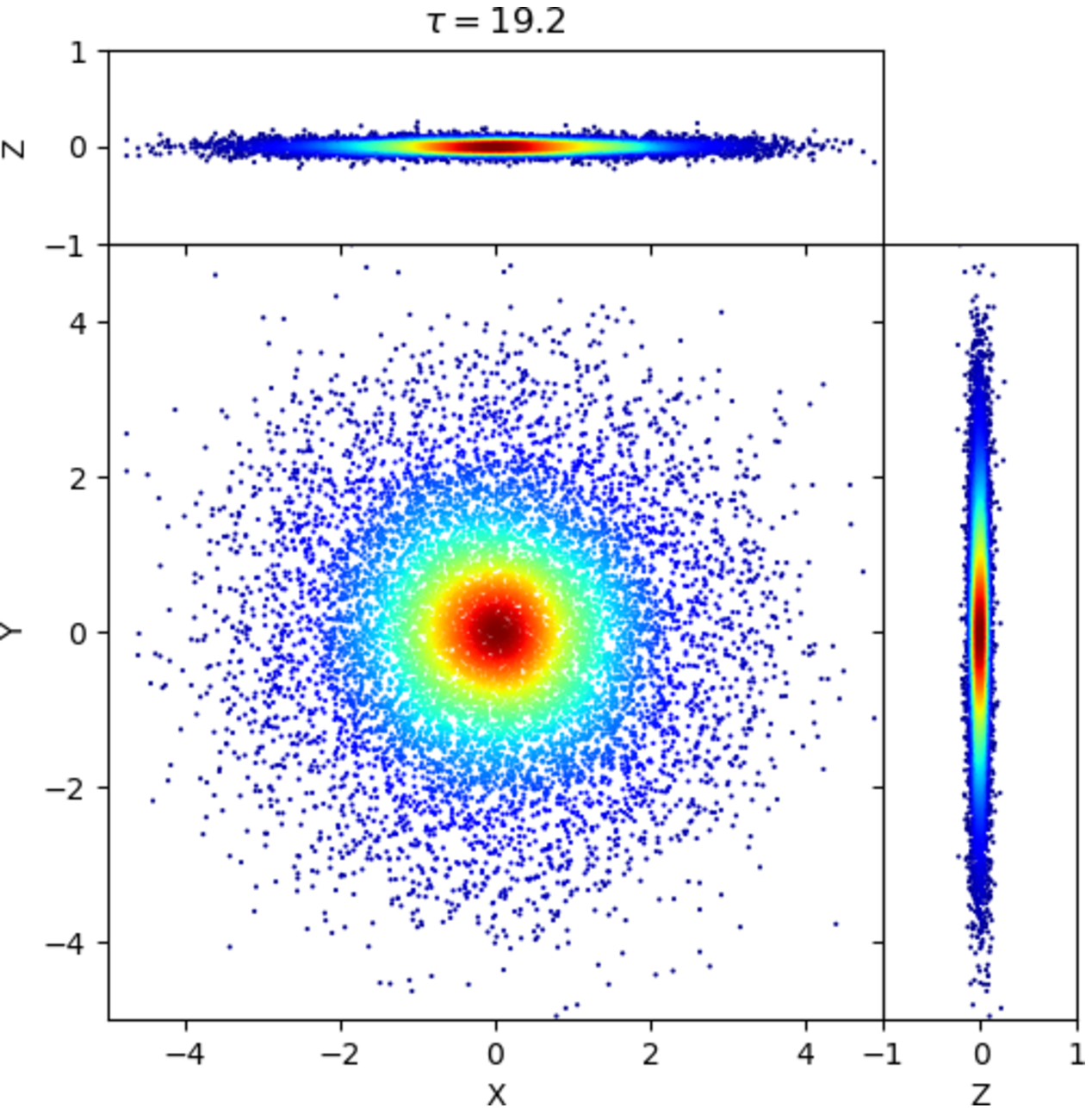}}\vspace{0.1 cm}
\centerline{\includegraphics[width=5.5cm]{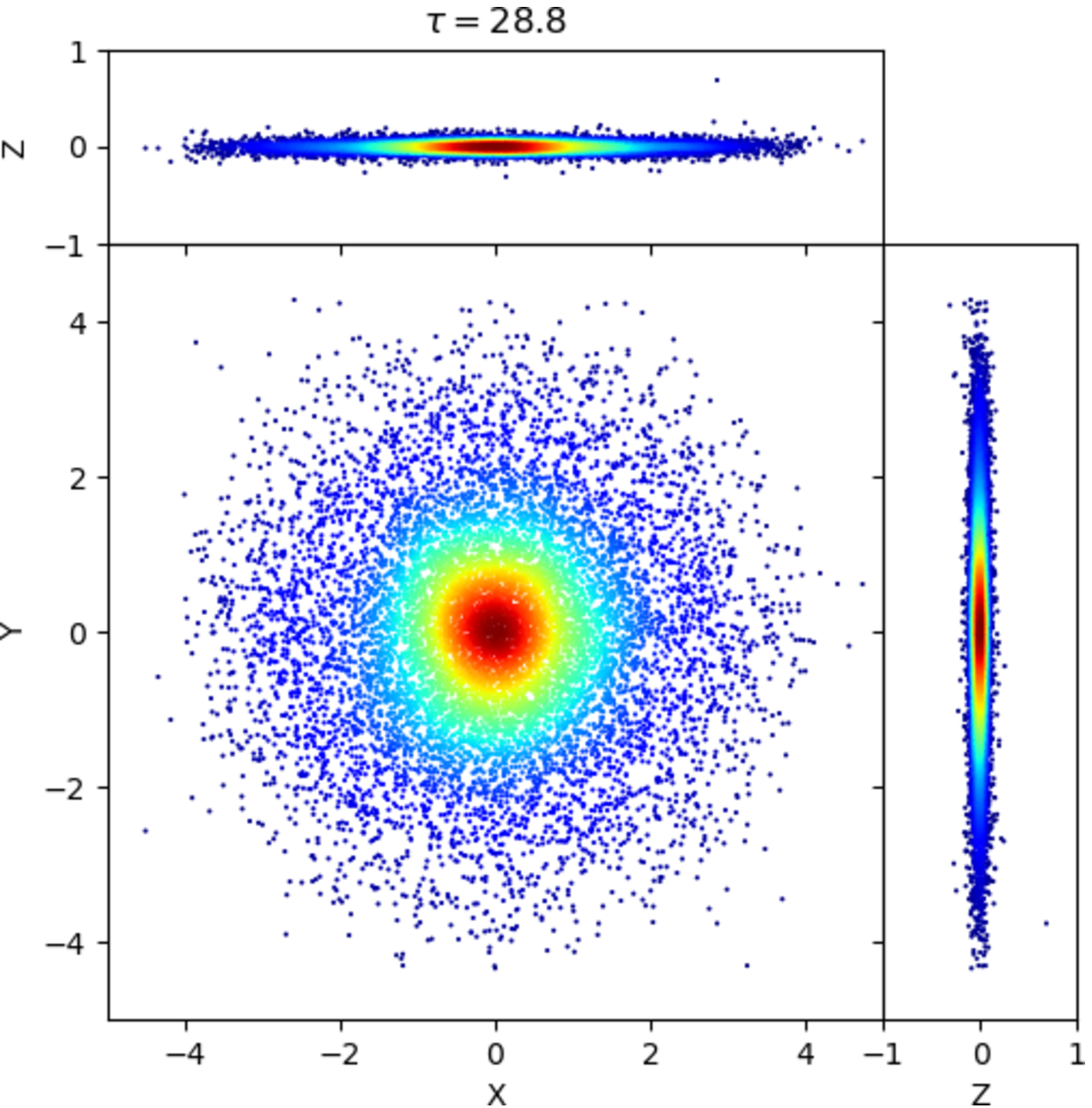} \hspace{0.01 cm}\includegraphics[width=5.5cm]{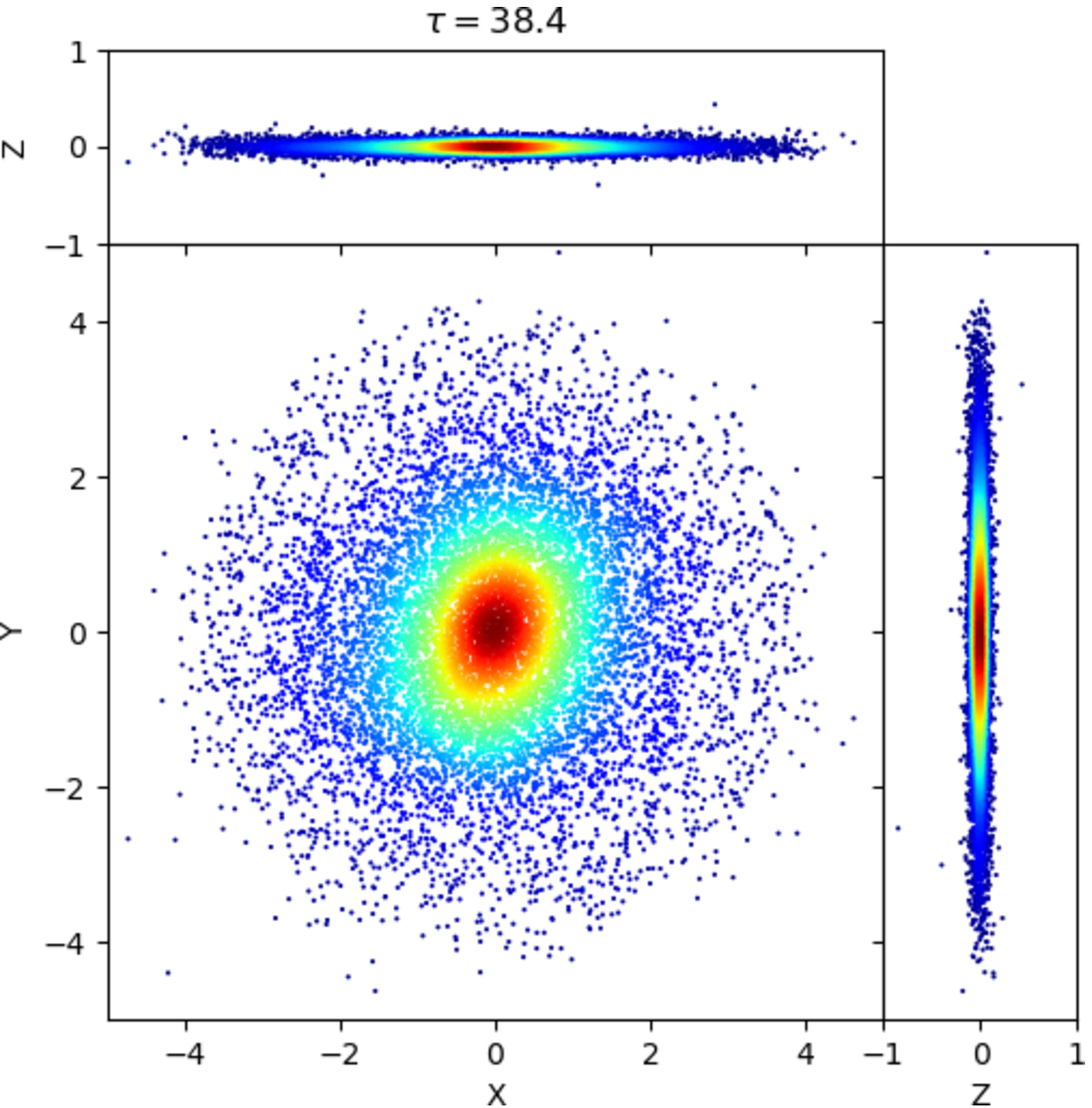}\hspace{0.01 cm} \includegraphics[width=5.5cm]{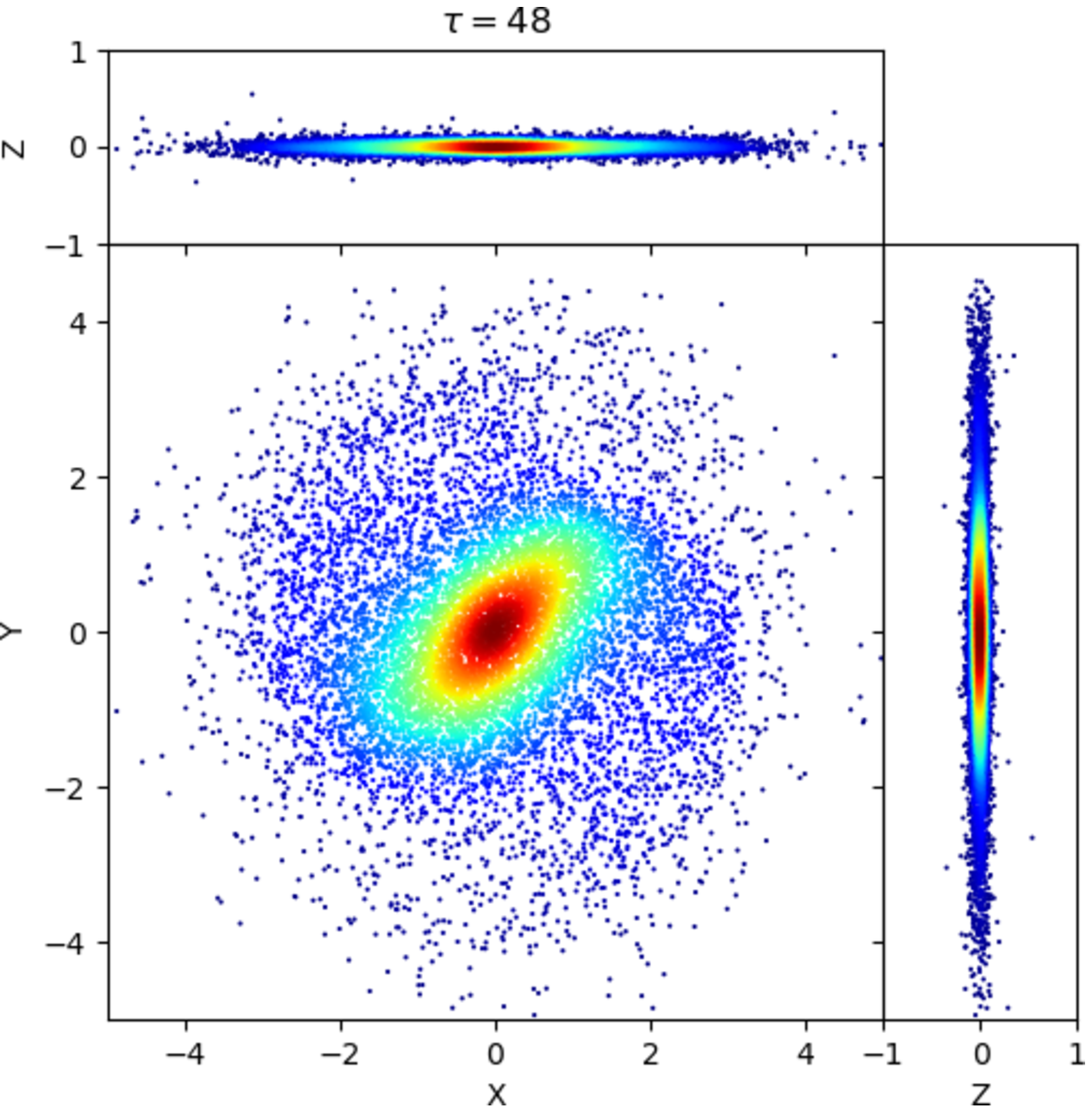}}\vspace{0.1 cm}
\centerline{\includegraphics[width=5.5cm]{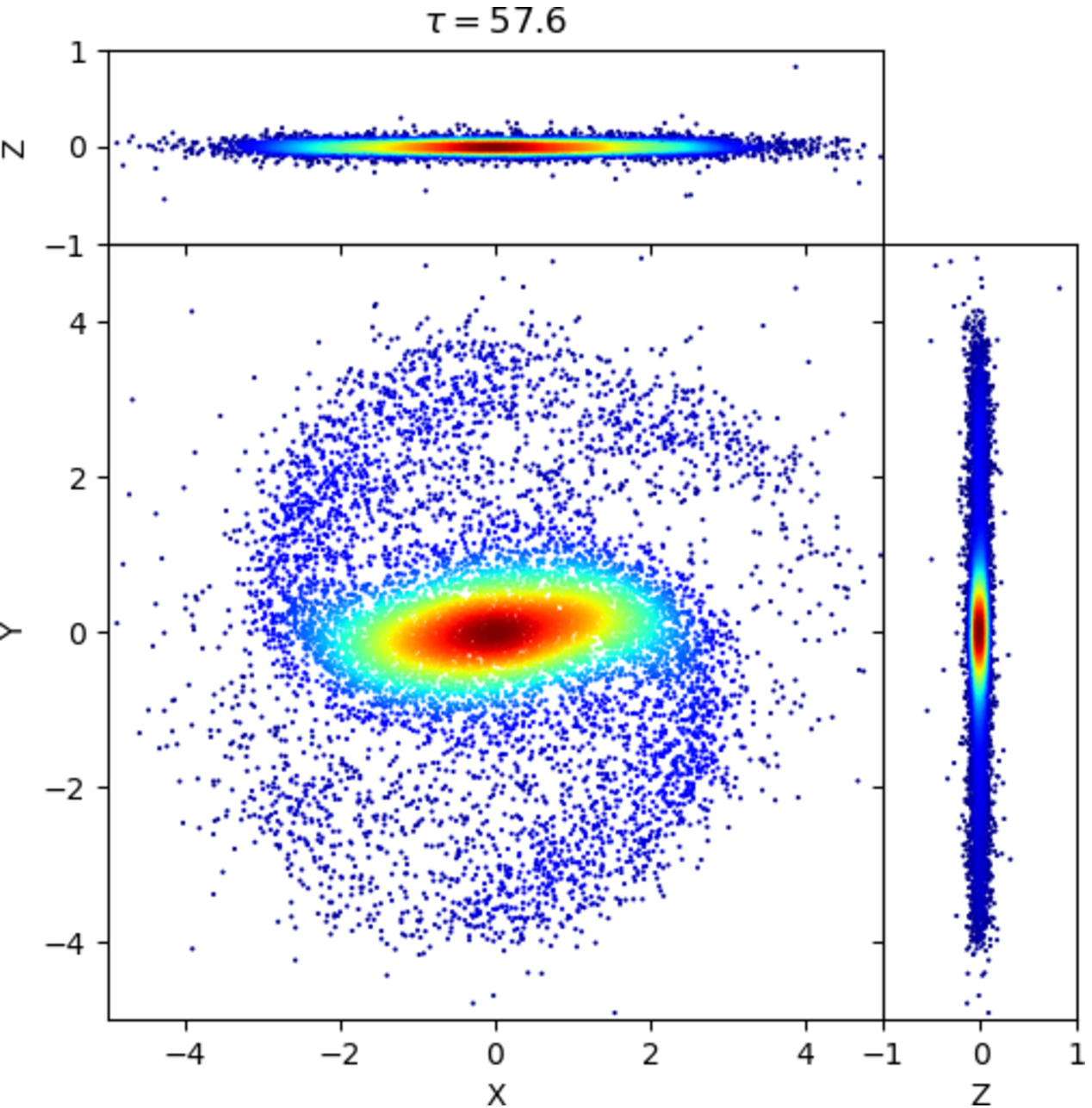}\hspace{0.01 cm} \includegraphics[width=5.5cm]{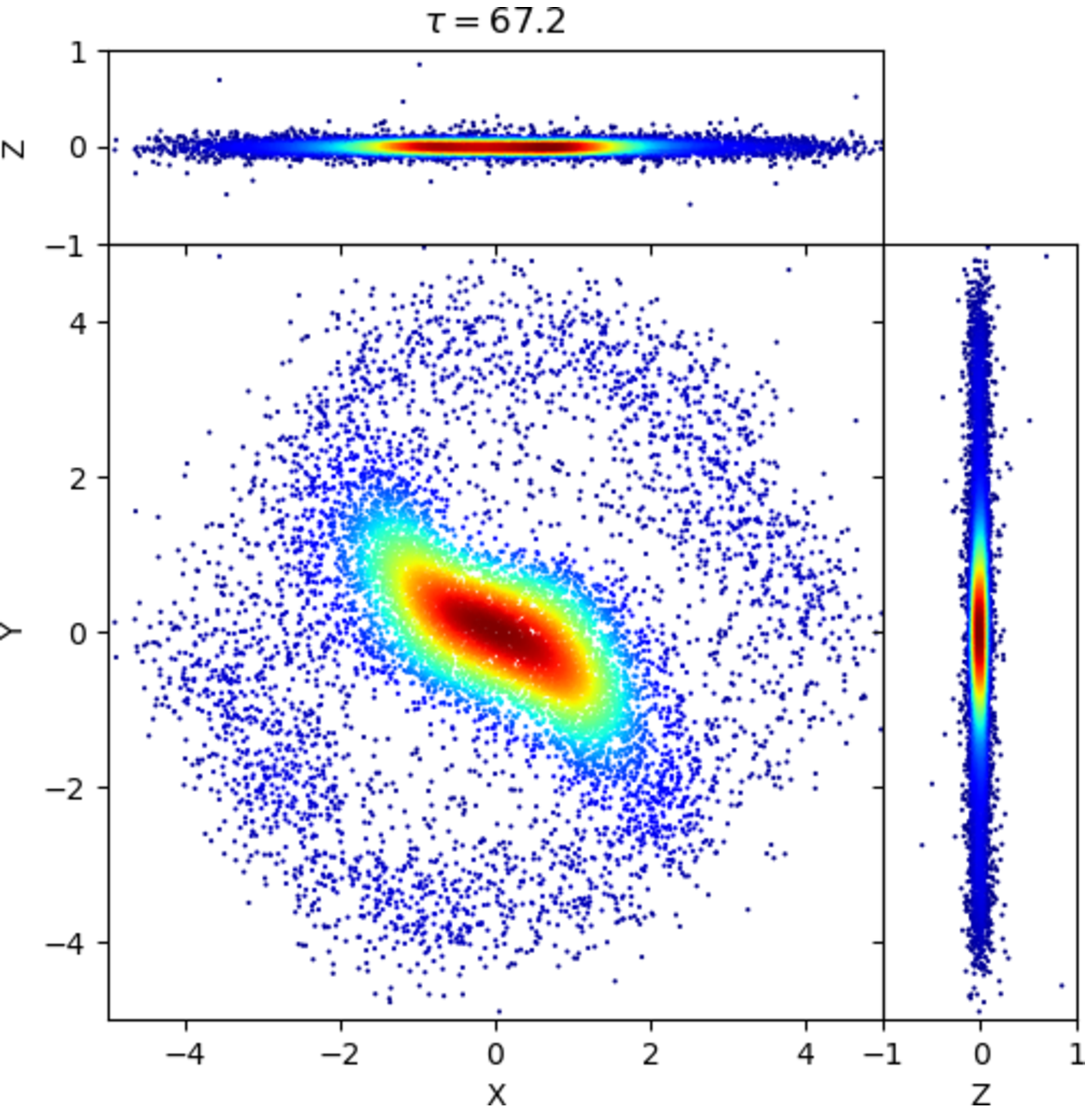}\hspace{0.01 cm}\includegraphics[width=5.5cm]{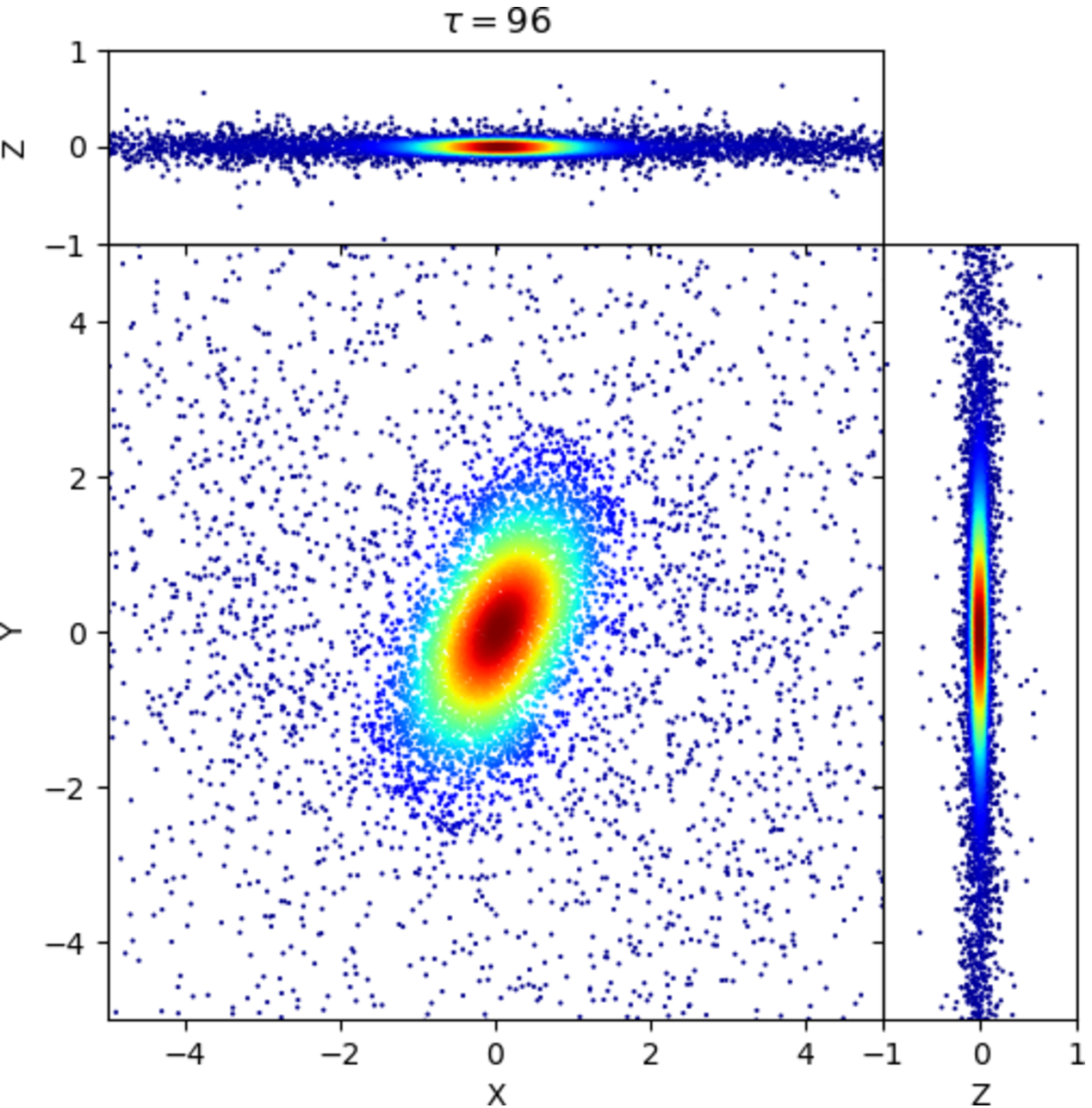}}
\caption{{The evolution of the disk, with $2\times 10^7$ particles in model ENLG, projected on the $x-y$, $x-z$ and $y-z$ planes, in the first part of the simulation. In this part, the bar strength reaches its maximum while the thickness of the disk remains small. For better visualization, only $8\times 10^3$ particles have been shown, and smoothed in central parts, in each panel.}}
\label{pos1}
\end{figure*}
 \begin{figure*}
\centerline{\includegraphics[width=5.5cm]{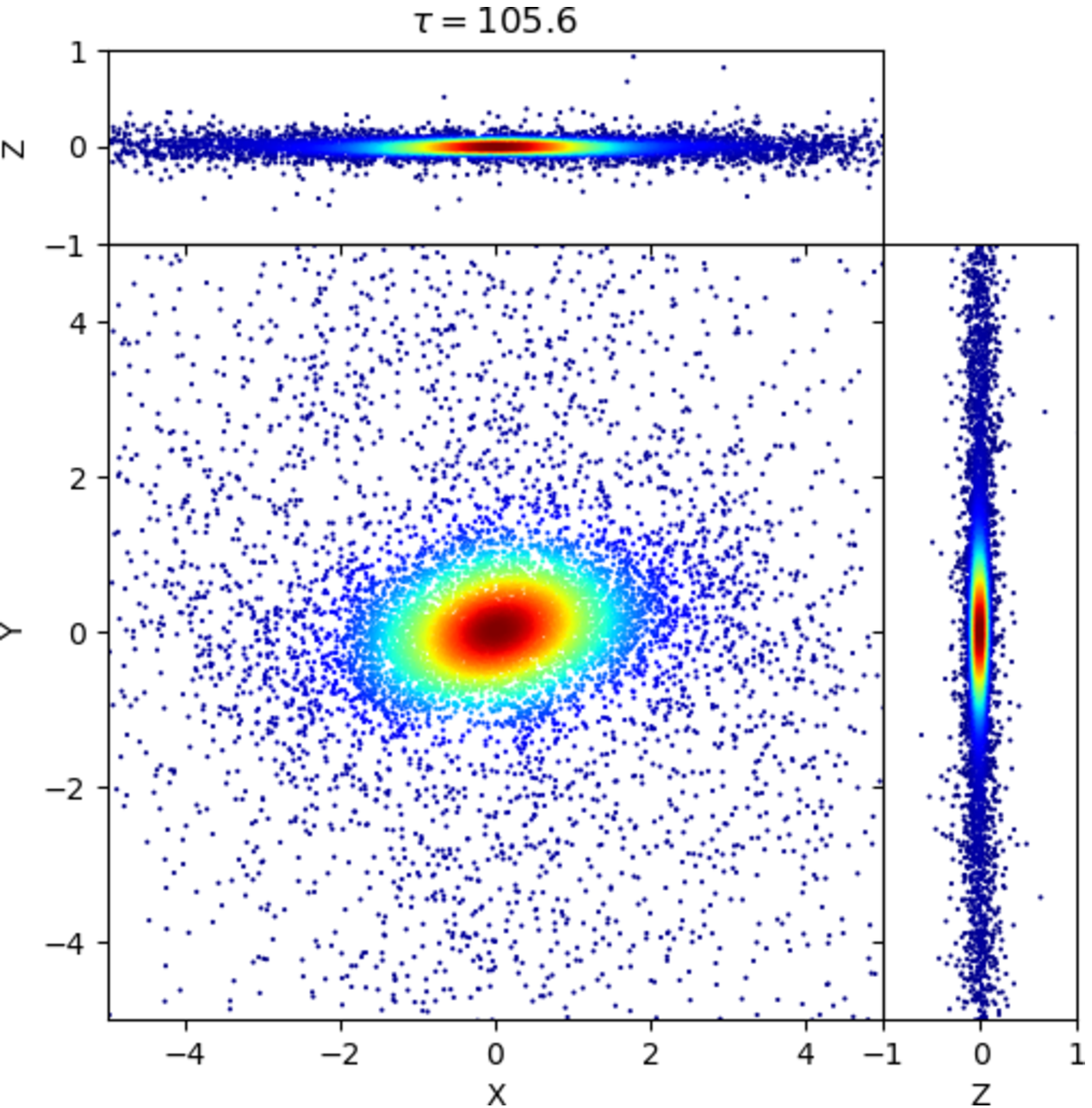}\hspace{0.01 cm} \includegraphics[width=5.5cm]{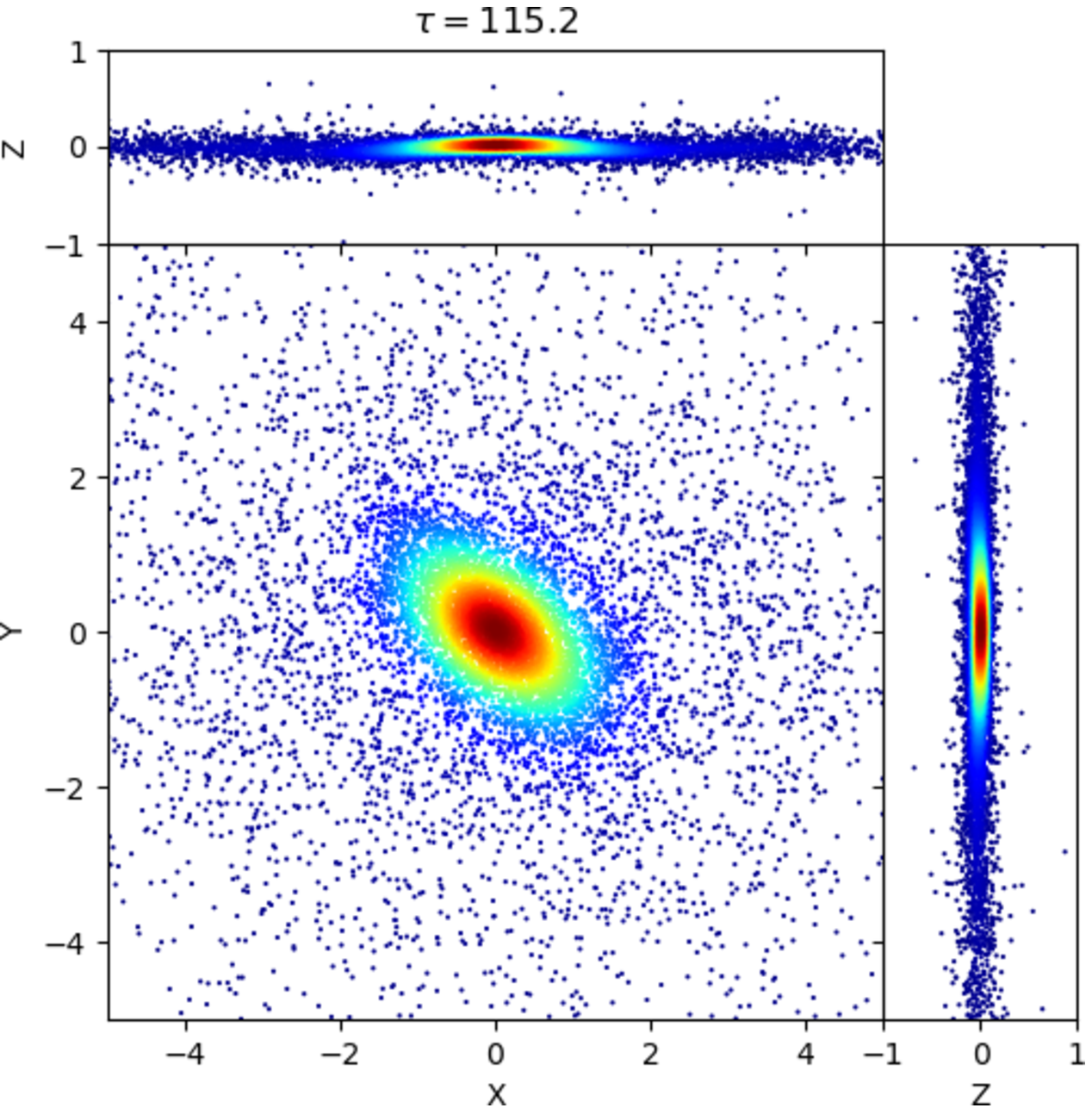}\hspace{0.01 cm} \includegraphics[width=5.5cm]{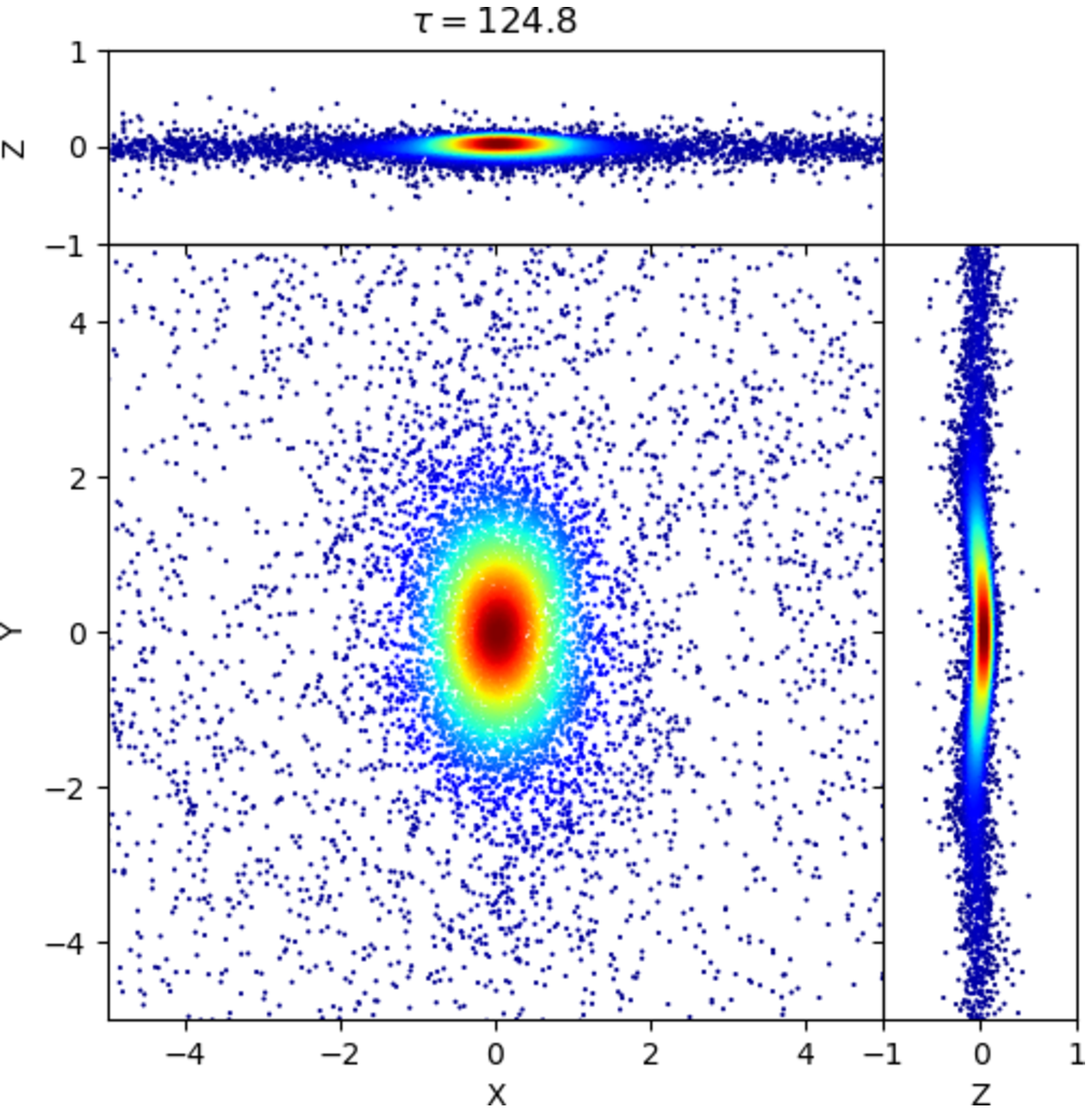}}\vspace{0.1 cm}
\centerline{\includegraphics[width=5.5cm]{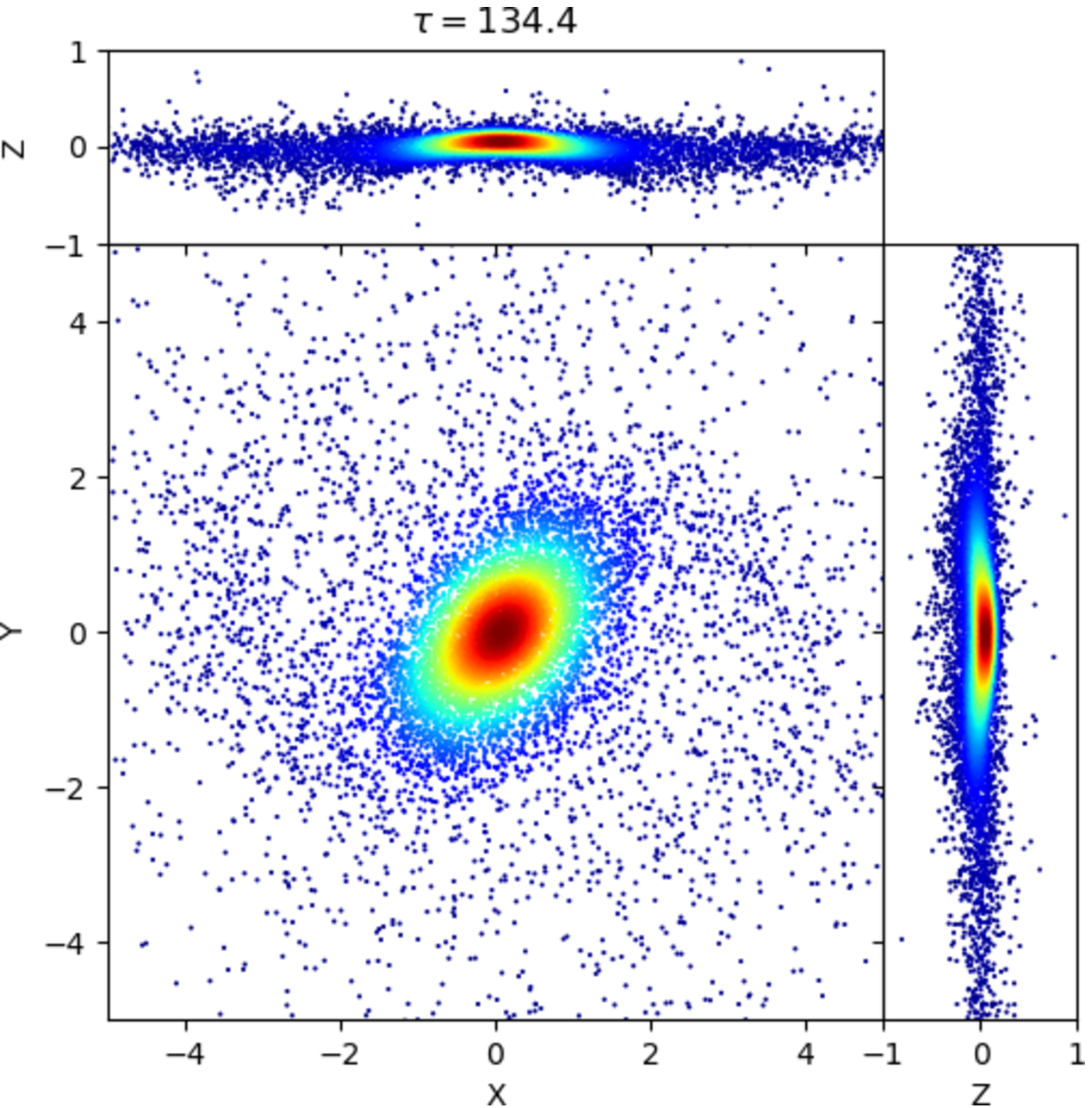} \hspace{0.01 cm}\includegraphics[width=5.5cm]{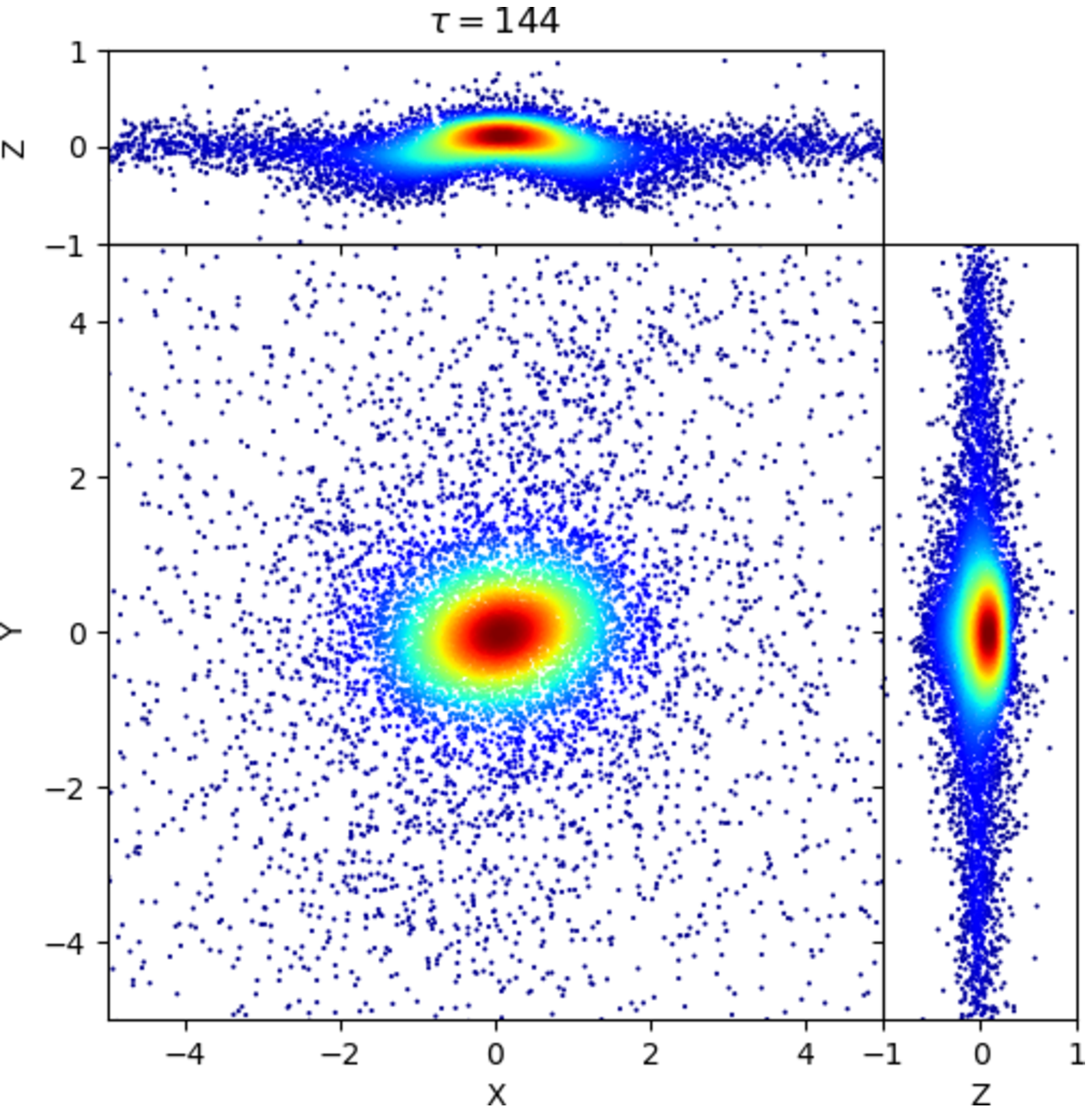}\hspace{0.01 cm} \includegraphics[width=5.5cm]{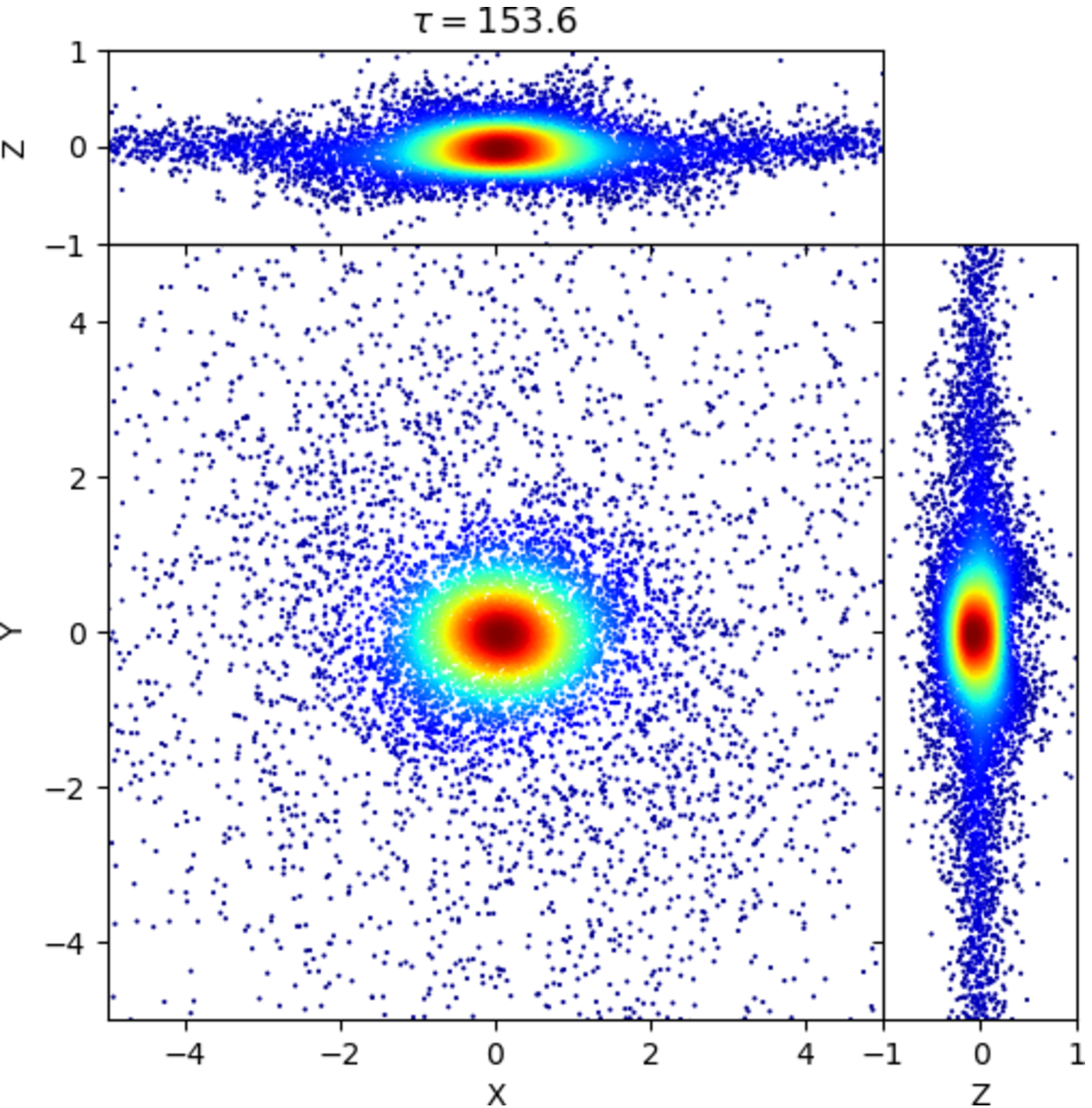}}\vspace{0.1 cm}
\centerline{\includegraphics[width=5.5cm]{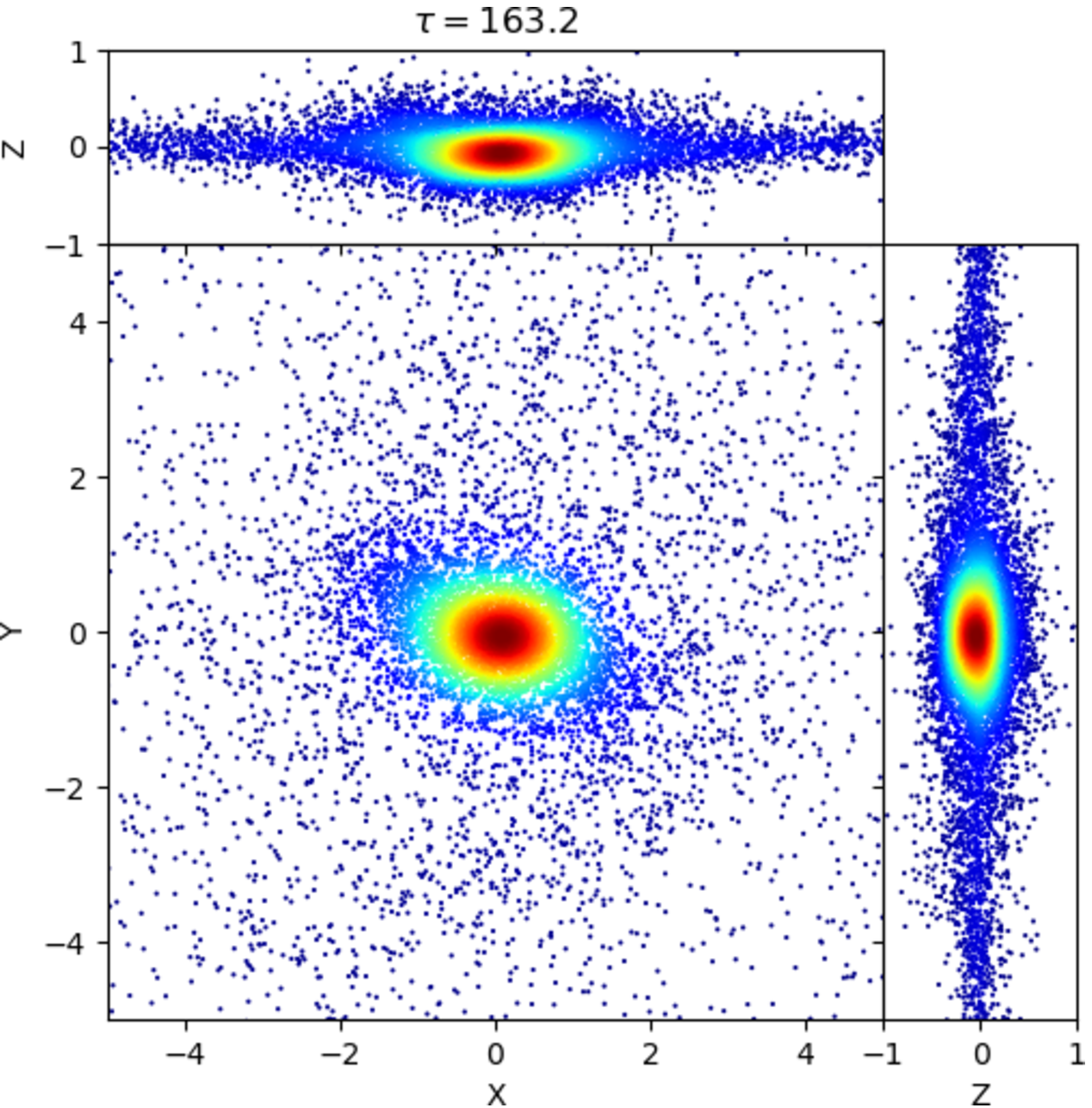}\hspace{0.01 cm} \includegraphics[width=5.5cm]{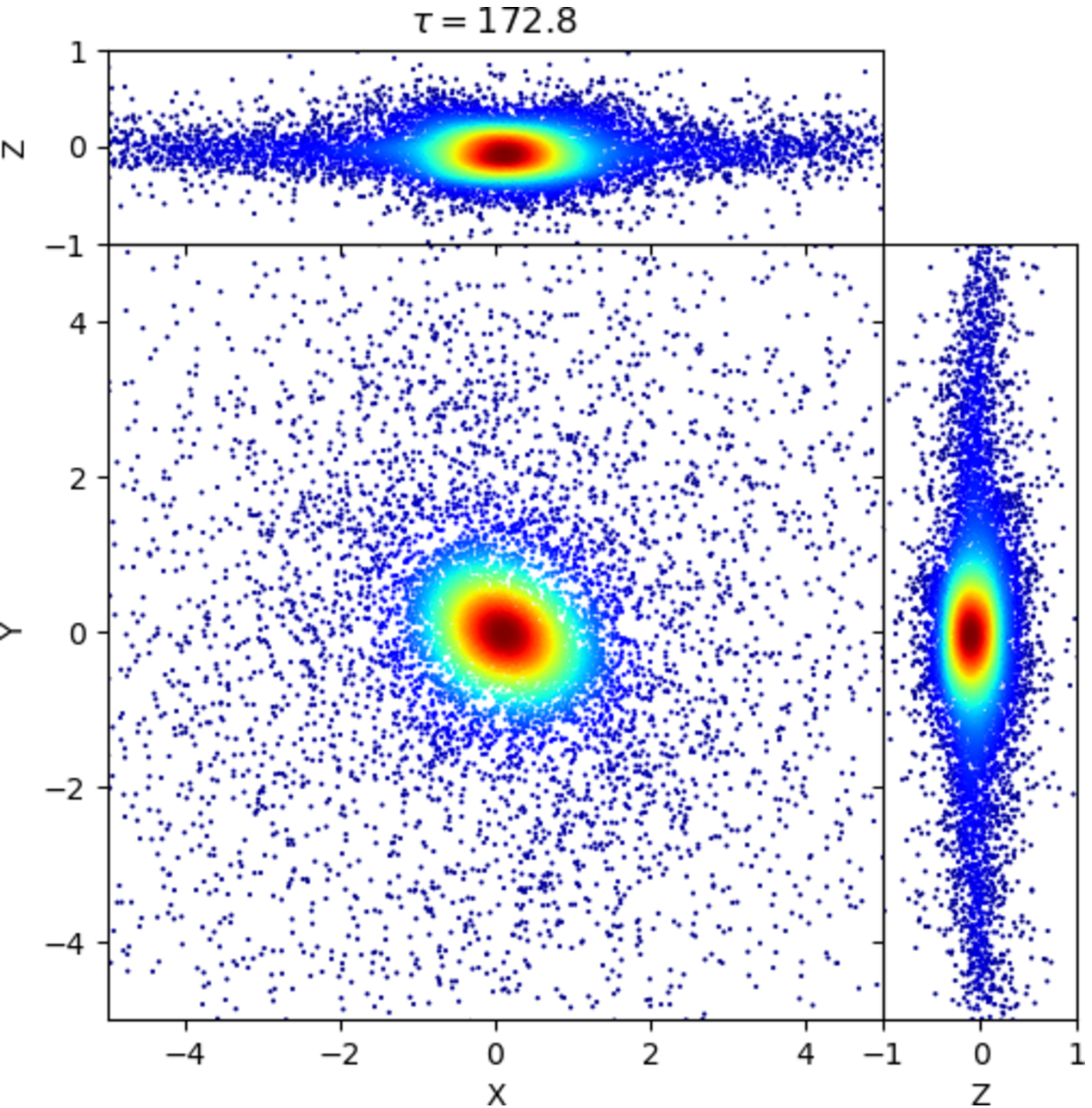}\hspace{0.01 cm}\includegraphics[width=5.5cm]{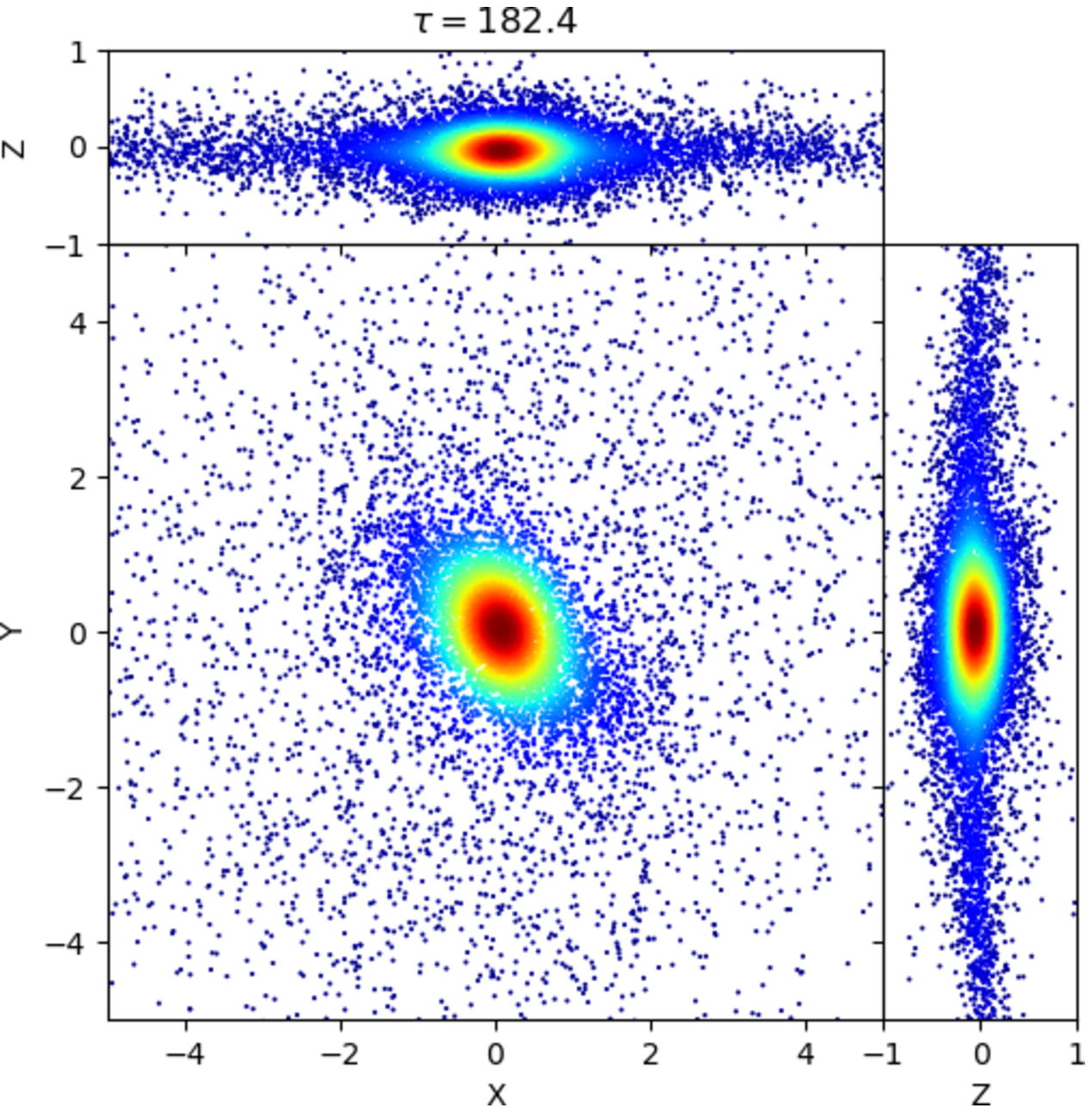}}
\caption{The evolution of the disk, with $2\times 10^7$ particles in model ENLG, projected on the $x-y$, $x-z$ and $y-z$ planes, in the second part of the simulation. In this part the bar strength reaches its minimum and the thickness of the disk starts to increase. For a better visualization, only $8\times 10^3$ particles have been shown, and smoothed in central parts.}
\label{pos2}
\end{figure*}
 
The amplitude of the non-axisymmetric features on the surface of the disk can be determined by the complex Fourier coefficient
\begin{equation}
A_m(\tau)=\frac{1}{N}\sum_{j=1}^{N} e^{i\,m(\phi_j(\tau)+\cot \gamma \ln r_j)}
\label{ex}
\end{equation}
where $(r_j , \phi_j )$ are the position of the
$j$th particle at time $t$ in the cylindrical polar coordinates. Moreover $\gamma$ is the pitch-angle of an $m$-fold logarithmic spiral. Logarithmic spirals make a complete basis set and consequently can be used to expand any two-dimensional function $f(r,\phi)$ as follows
\begin{equation}
A_m(\tau)=\int_0^{\infty}\int_0^{2\pi}f(r,\phi)e^{i\,m(\phi(\tau)+\cot \gamma \ln r)}r \,dr\, d\phi
\label{ex2}
\end{equation}
here $f$ is the surface density and is given by a set of delta functions. Therefore one may verify the Fourier coefficient \eqref{ex}. Notice that the summation is over the disk particles and does not include halo particles in the model EPL. The bar amplitude, for which $m=2$, is computed by the ratio $A_2/A_0$. For simplicity, we assume that $\cot \gamma=0$. In this case, $A_2$ contains both bar-like and spiral contributions. However, as we shall see, the spiral pattern exists in a short period of time compared with the time duration of the simulations, and we mostly deal with bar-like structures.
 \begin{figure*}
\centerline{ \includegraphics[width=17cm]{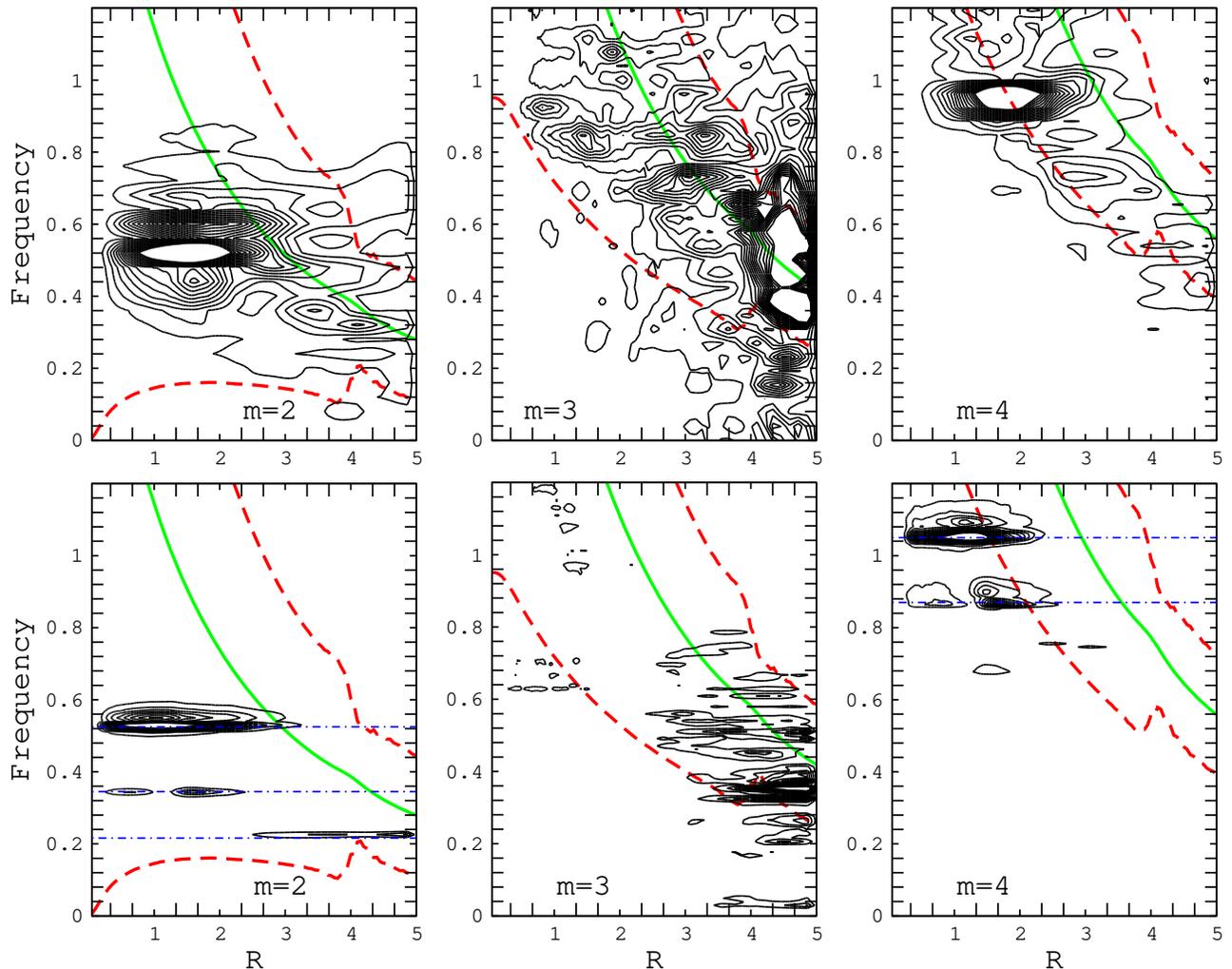}}
\caption{{Isocontours of the power spectrum as functions of $r$ and $\omega$ for different sectoral harmonics $m$ in model ENLG with $2\times 10^7$ particles. The top row covers the first part of the evolution, i.e., $10 \leq \tau \leq 160$, and the bottom row belongs to the second part of the simulation, namely $160<\tau<800$. In each panel, the green curve indicates $m \Omega_c$ and the dashed red curves are $m \Omega_c \pm \kappa$. Each horizontal ridge shows the existence of a coherent density wave with a well-defined angular frequency that can be simply identified from the vertical axis.}}
\label{power}
\end{figure*} 

The evolution of the bar in our models, has been shown in Fig. \ref{bar}. The blue curve belongs to our main model ENLG, the solid black curves are the corresponding curve for the model EPL. Furthermore, for comparison, the evolution of the bar for the model EM, has been illustrated with the black dashed curve. It is clear that the bar instability develops more rapidly in NLG compared with the standard dark matter model. More specifically, the exponential growth rate, i.e., $e^{\omega \tau}$, obtained from the top panel in Fig \ref{bar}, in ENLG, EPL and EM models are $\omega\simeq 0.097$, $\omega\simeq 0.067$ and $\omega\simeq 0.058$ respectively. This means that growth rate in model ENLG is $67$\% ($45$\%) larger than that of the model EM (EPL). On the other hand we see that MOG has stabilizing effect, for more details on the stabilizing effects of MOG we refer the reader to \cite{gr2017} and \cite{roshan2018}. It is interesting that although both theories, i.e., NLG and MOG, add almost similar Yukawa corrections to the gravitational force in the weak field limit, the onset of the bar instability and its evolution are quite different.

The rapid bar growth in NLG is reminiscent of the rapid growth in MOND. It is shown in \cite{ti} that bar grows faster in MOND compared with the standard case. However, in MOND the maximum bar amplitude remains almost constant for a relatively long time interval ($4$ Gyr) and then decreases rapidly and stays almost constant at a much lower value, see Fig. 7 in \cite{ti}. On the other hand, in NLG the bar magnitude starts to increase at $\tau\simeq 10$, and reaches its maximum at $\tau\simeq 60$, see Fig. \ref{bar}. This  point is also clear from Fig. \ref{pos1}. In this figure we have projected the position of a small fraction of the particles on the Cartesian planes. This figure shows the evolution of the particles in the first part of the simulation where the thickness of the disk remains small. Furthermore, for better visualization the points have been smoothed, especially in the central part of the disk, using the Python module \texttt{scipy.stats.gaussian\_kde}. Naturally, from dark red to blue color, the surface density decreases. This smoothed scatter plots are helpful to discriminate and track the evolution of the bar in a visual way. We see that at $\tau\simeq 57.6$ a spiral density wave is excited and rapidly fades its arms. 

It is important to mention that in the EPL model, the bar magnitude experiences a relatively rapid growth at later times. In fact, the Newtonian bar can strengthen again (after the main drop), by exchanging angular momentum with the dark halo \cite{at2002}.

As it is clear from Fig. \ref{bar}, unlike in MOND, the bar magnitude starts to decrease and reaches its minimum value at $\tau\simeq 150$. We see that after this minimum, a new phase in the evolution is started and the bar grows with a very small slope. The second part of the evolution, where the bar's magnitude is small and the thickness of the disk starts to increase, has been illustrated in Fig. \ref{pos2}. It should be noted that there are two main features apparent in figures \ref{bar} and \ref{pos2}. We see that the bar strength oscillates with an almost constant frequency in the second part of the simulation. Finding the power spectra of the model is helpful for revealing the reason for this oscillation. On the other hand it is clear from Fig. \ref{pos2} that the thickness of the disk suddenly starts to increase with time, and it seems that the mean thickness, i.e., $\langle z\rangle$, oscillates around $z=0$ for a short period of time. In the following, we study these both cases with more detail.

\subsection{Power spectra of the model ENLG}
Let us start with the rapid oscillations in the magnitude of the bar. One should note that, as we already mentioned, the bar magnitude $A_2$ contains the contributions from spiral structures and the bar-like pattern as well. Consequently, this oscillation can be interpreted as the existence of beating/coupling between different modes. It is interesting that this oscillation exists only in modified gravity models, i.e., in the models ENLG and EM.

For revealing the existence of different modes, one may naturally find the power spectrum of the mass distribution. To do so, at each output time we first perform a one-dimensional Fourier transform of the surface density in $\phi$ (notice that the transform \eqref{ex2} is two-dimensional and taken over $r$ and $\phi$). The result for each mode $m$ will be a function of $r$ and $t$ as $A_m(r,t)$. In order to find the power in terms of $r$ and frequency $\omega$ for each sectoral harmonic $m$, we calculate the Fourier transformation in time of the coefficients $A_m(r,t)$ and show it by $\tilde{A}_{m}(r,\omega)$. In this case the power spectrum is given by $\vert\tilde{A}_{m}(r,\omega)\vert^2$.

The isocontours of the power spectrum for our main model ENLG has been shown in Fig. \ref{power}. We have shown power spectra for three different modes $m=2$, $3$ and $4$, where there are $2\times 10^7$ particles in the simulation. Furthermore, considering the overall behavior of bar magnitude shown in Fig. \ref{bar}, we have taken the temporal Fourier transformation over two different time intervals. The top row in Fig. \ref{power} is for the first part of the evolution $10<\tau<160$, and the bottom row is for the second part of the evolution, i.e., $160<\tau<800$.

In this figure, the solid (green) curves mark $m \Omega_c$ and the dashed (red) curves show $m \Omega_c \pm \kappa$. Where $\Omega_c$ and $\kappa$ denote the angular frequency for the circular motion and the Linblad epicycle frequency respectively. Notice that instead of theoretical curves, we have computed these frequencies using the initial velocities at $\tau=0$. Subsequently, one can see a rapid change in the red curves at $r\simeq 4$, which is related to the surface mass density truncating procedure explained in the section \ref{gi}. 

It is clear from the top row that, in the first part of the simulation there are two bar-like modes inside the corotation radius. Notice that each horizontal ridge determines a coherent density wave. In fact these $m=2$ modes, with close frequencies $\omega\simeq 0.6$ and $\omega\simeq 0.52$, start from the inner parts of the disk and stop at the corotation. One should note that the oscillations in the bar magnitude cannot be a consequence of beating between these two modes. The beat period, in this case, is large and does not coincide the period that can be inferred from Fig. \ref{bar} ($\tau_b\simeq 20$). Also there is a major contribution from the $m=4$ mode with frequency $\omega\simeq 0.94$. The $m=3$ mode has not an important impact on the system. In fact by plotting its magnitude it turns out that $A_3/A_2$ and $A_3/A_4$ remain small.

In the second part of the simulation, there are three $m=2$ modes inside the corotation, see bottom row in Fig. \ref{power}. We have shown them with blue dot dashed straight lines at $\omega_1\simeq 0.22$, $\omega_2\simeq 0.35$ and $\omega_3\simeq 0.52$. The last one is the long-lived dominant mode which was excited from the beginning of the simulation. On the other hand, we see that there are also two $m=4$ density waves with frequencies $\omega_5\simeq 0.87$ and $\omega_6\simeq 1.5$. The existence of these five waves is the main reason for the rapid oscillation in bar amplitude. Since we deal with several modes, analysis of the coupling and beating between the modes is somehow complicated. However, it seems that the beating between modes $\omega_1$ and $\omega_3$, which yields a beating frequency $\omega_b=\omega_3-\omega_1$ and period $\tau_b\simeq 20.9$, is responsible for the oscillations. This frequency is consistent with that of obtained from Fig. \ref{bar}.

As our final remark in this subsection, it is interesting to mention that two bar-like beat waves can appear as an $m=4$ mode. The frequency of the resultant $m=4$ mode with frequency $\omega_5$, can be obtained by the sum of the frequencies of $m=2$ modes, see \cite{masset} for more details. From this perspective, we see that the slower $m=4$ mode can be considered as the outcome of an interaction between bar-like modes with $\omega_3\simeq 0.52$ and $\omega_2\simeq 0.35$.

  \begin{figure} 
\centerline{\includegraphics[width=8.5cm]{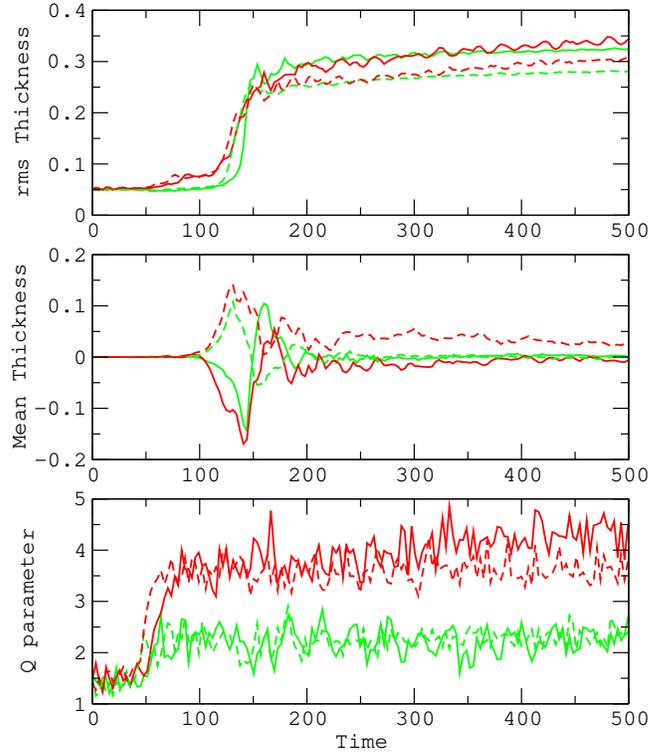}}
\caption{The top panel shows the evolution of the rms thickness of the disk with respect to time. The middle panel is the mean thickness of the disk with respect to time. In the bottom panel, the Toomre parameter $Q$ is shown as a function of time. In all the panels, the solid and dashed curves belong to simulations with $N=2\times 10^7$ and $3\times 10^6$ particles respectively. On the other hand, red and green colors belong to quantities at $R=1.1$ and $R=2.1$ respectively.}
\label{peanut}
\end{figure}
\subsection{Vertical structure of the model ENLG}
Now let us return to the vertical structure of the disk in the ENLG model. As we already mentioned, it is clear from Figs. \ref{pos1} and \ref{pos2}, that there is a bulk oscillation around $z=0$ plane. Furthermore the rms thickness $\sqrt{\langle z^2 \rangle}$ of the disk suddenly starts to increase at $\tau\simeq 150$. To see this point in a more quantitative way, we have plotted the time evolution of rms thickness at two different radii in the top panel in Fig. \ref{peanut}. More specifically, the solid and dashed curves belong to the model with $N=2\times 10^7$ and $N=3\times 10^6$ respectively. Moreover, in all the panels of Fig. \ref{peanut} green and red colors belong to quantities at $R=1.1$ and $R=2.1$ respectively. It is clear that, as expected, there is no significant difference between simulations with different $N$. 

In agreement with Fig. \ref{pos2}, we see that that $\sqrt{\langle z^2 \rangle}$ has a rapid and step-like behavior around $\tau=150$. More specifically during the buckling, $\sqrt{\langle z^2 \rangle}$ increases from $0.05\, R_d$ to $0.3\, R_d$. From the middle panel, it is seen that at the same time $\langle z \rangle$ oscillates around the plane of the initial disk. It is interesting to mention that the rapid rms thickness growth coincides with a substantial decrease in the bar magnitude. This behavior can be simply interpreted by considering the fact that particles resonate in the $z$ direction. In other words, the vertical oscillation can be amplified to form a peanut shape \citep{cs}. On the other hand, we know that the primary effect of the thickness is to reduce the growth rate of the Jeans instability in the system, see \cite{romeo1992stability}. Consequently, it is natural to expect a reduction in the bar magnitude through the stabilizing effects rooted in the thickening of the disk.

The Toomre's parameter is another quantity directly related to the stability of the disk. As already mentioned this parameter accounts for the local stability of the system. To be specific, the disk is locally stable against small perturbations provided that $Q>1$. On the other hand, it is well-known that the \textit{swing amplification} is extremely sensitive to the value of $Q$. When $Q$ is close to unity, this mechanism works effectively to form $m=2$ patterns. However, the amplification factor decreases rapidly with increasing $Q$ \citep{julian}. It also provides additional insight into the heating rate due to the activity of the bar. In the bottom panel of Fig. \ref{peanut}, we have plotted the evolution of the $Q$ parameter in two different radii. It is clear that $Q$ starts to increase when the bar is developed in the disk at $\tau\simeq 50$. This directly means that the occurrence of the bar instability heats the disk. Moreover, the bar influences the larger distances more effectively than the inner radii. After $\tau\simeq 75$ the parameter $Q$ gets almost constant in both radii as $Q(1.1)\simeq 2.25$ and $Q(2.1)\simeq 3.75$. It is interesting that the buckling instability which happens at $\tau \simeq 150$ does not have any impact on the $Q$ parameter. In other words, $Q$ increases due to the bar activity and not thickening of the disk by buckling instability. This directly means that the bar instability seriously heats the disk, while the short-term asymmetry around the disk plane caused by the buckling instability does not. The relatively large value for $Q$ along with the buckling mentioned above prevents further bar growth. The behavior of $Q$ with respect to time in the ENLG model is similar to what happens in MOND. It is shown in Fig. 17 of \cite{ti} that $Q$ rises in MOND and then gets almost constant. However, it grows continuously in the dark matter model.

Comparing the vertical structure of the model ENLG with the models EPL and EM  studied in \cite{roshan2018}, one may conclude that the final rms thickness of the disks in NLG is similar to those of standard dark matter model (EPL). On the other hand, the model in MOG has a significant difference in the sense that its final rms thickness is substantially smaller than in the dark matter mode. In other words, our simulation shows that, unlike in MOG and MOND which lead to stronger peanuts \cite{ti}, the model in NLG produces similar peanuts compared with the model in standard dark matter paradigm.

 \begin{figure}
\centerline{\includegraphics[width=8.5cm]{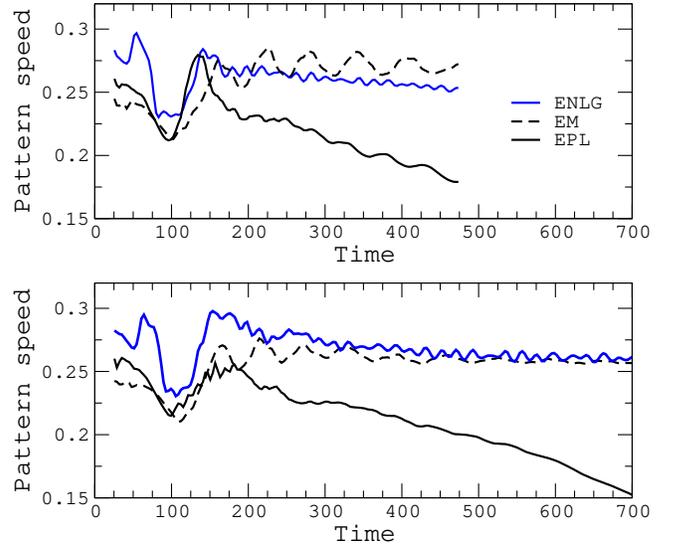}}
\caption{The pattern speed $\Omega_p(\tau)$ with respect to time. The blue curve belongs to our main model ENLG, the dashed curves are for the model in MOG, and the solid black curves belong to dark matter model EPL. The latter two cases are shown just for comparison. In the top panel there are $3\times 10^6$ particles in each live component, and in the bottom panel, there are $2\times 10^7$ particles.}
\label{ps}
\end{figure}

\subsection{Bar's pattern speed $\Omega_p$}
Beside the bar growth in NLG, the other important quantity which is also related to the dark matter problem is the bar's pattern speed $\Omega_p(\tau)$. In this section, we put emphasis on the crucial significance of the pattern speed's evolution in order to discriminate between dark matter particles and modified gravity. We have measured $\Omega_p$ in terms of time and the result has been illustrated in Fig. \ref{ps}. The top and bottom panels show the evolution of $\Omega_p$ for $N=3\times 10^6$ and $N=2\times 10^7$ particles respectively. The blue curve belongs to the model ENLG. On the other hand solid and dashed black curves are corresponding curves in EPL and EM models.
\begin{figure*} 
\centerline{\includegraphics[width=12cm]{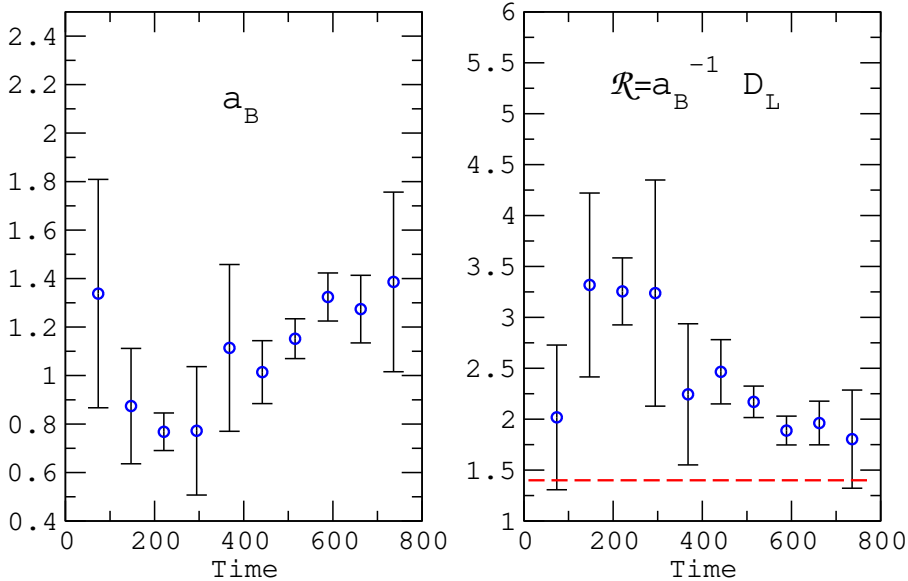}}
\caption{The ratio $\mathcal{R}=D_L/a_B$ for the model ENLG with $2\times 10^7$ particles in each live component, calculated at some different times. The corotation radius is almost constant at $D_L\simeq 2.5$. The error bars are obtained from the averaging procedure introduced for calculating the bar length. The dashed red line indicates $\mathcal{R}=1.4$ as the border to separate the fast and slow bars. }
\label{ratio}
\end{figure*}

As expected the pattern speed slows down with time in the EPL model. The dynamical friction between halo and disk particles is the main reason for this reduction. In our main model, i.e., ENLG, after $\tau\simeq 150$ the bar magnitude gets almost constant and we can assign a well defined pattern speed. As we see in Fig. \ref{ps}, the pattern speed starts to decrease with a small slope. The small amplitude oscillation in the pattern speed is a result of the beating between different modes. One can easily verify that this oscillation's frequency is consistent with the beating frequency discussed in the previous section. Since $\Omega_p$ is almost constant in ENLG, we can say that dynamical friction between disk particles is not effective in this case. This is also the case in the EM model shown with dashed curves in Fig. \ref{ps}. It is interesting that the dynamical friction does not appear in MOND as well \cite{ti}. It should be noted that the dynamical friction not only is important in the dynamics of isolated galaxies but plays role in the interaction between galaxies. For example, it is a key mechanism to increase the merger frequency of spiral galaxies during the cosmic evolution \cite{merge}. Therefore, this issue deserves more precise investigations to reveal observable differences between standard dark matter paradigm and modified gravity.

The relevance and importance of the pattern speed in modified gravity theories get more obvious when almost all the current observations show that spiral galaxies host fast bars \cite{pattern}. However, dark matter simulations for both isolated galaxies and cosmological simulations in $\Lambda$CDM predict an opposite result. More specifically, these simulations show that, mainly due to the dynamical friction, simulated spiral galaxies host slow bars, for a pioneer paper that pointed out this challenge in isolated galaxies see \cite{deb2000}. Furthermore for a recent cosmological simulation that also confirms this discrepancy see \cite{cosmo}.

Therefore it seems necessary to compare our simulation with the pattern speed data. Notice that $\Omega_p$ cannot directly specify that whether the bar is fast or slow. Instead, it is convenient to use the ratio of the corotation radius to the bar semi-major axis $\mathcal{R}$ defined as
\begin{equation}
\mathcal{R}=\frac{D_L}{a_B}
\label{m9}
\end{equation}
where $D_L$ is the corotation radius and can be obtained by measuring the pattern speed $\Omega_p$. On the other hand, $a_B$ is the bar semi-major axis. In this case the bar is fast if $\mathcal{R}\lesssim 1.4 $, and is slow if $\mathcal{R} \gtrsim 1.4$. At each time step $\tau$, we measure the pattern speed $\Omega_p$ and the angular velocity $\Omega(R)$. Therefore it is straightforward to find the coratation radius by solving $\Omega(D_L)=\Omega_p$. As we see in Fig. \ref{ps}, the pattern speed is almost constant in the ENLG model. Accordingly, the corotation radius $D_L$ does not move throughout the disk. Therefore the time variation of $a_B$ specifies the evolution of the $\mathcal{R}$ parameter.

Now our main task is to measure the bar semi-major axis. To measure $a_B$ in terms of time, we use a simple procedure already explained in \cite{roshan2018}. At a given time $\tau$, working in the projected $x-y$ plane, we find a straight line parallel to the bar which crosses bar's center. This line can be simply expressed as $y = \tan \phi(\tau) x$, where $\phi(\tau)$ is the angular displacement of the bar. Then we assume a rectangle with width $\Delta L$ and length $5$ around this line. Finally, the rectangle is divided into small elements with equal area, and the surface density at each element is calculated. We choose the bar length to be the radius at which the density of the particles is less than $B$ percent of the central element. We vary $\Delta L$ between $0.2$ and $1$, and choose three values for $B$, i.e., $10$, $20$ and $30$. Naturally, these variations cause an error bar in the final value of $a_B$ and consequently in $\mathcal{R}$ at each time.

The result has been illustrated in Fig. \ref{ratio} for the model ENLG. In the left panel, we have shown the bar semi-major axis $a_B$. We see that bar starts with a relatively large length and then decreases to reach a minimum at $\tau\simeq 200$ right after the buckling instability. Then $a_B$ starts to grow and this growth continues to the end of the simulation. On the other hand, we have plotted $\mathcal{R}$ in the right panel. We see that due to the existence of a minimum in $a_B$, there is a maximum in the time evolution of $\mathcal{R}$. In other words, the growth of the bar length together with the constancy of the corotation radius enforces $\mathcal{R}$ to decrease toward the border $\mathcal{R}=1.4$. Notice that observed bars lie in the interval $0.9\lesssim \mathcal{R}\lesssim 1.4$.

It should be stressed that in the EPL model, dynamical friction reduces the pattern speed and consequently the corotation radius increases with time. Therefore $\mathcal{R}$ is an increasing function of time and reaches $\mathcal{R}\simeq 4$ at the end of the simulation, see \cite{roshan2018}. On the other hand, the model EM in MOG has a similar behavior as in ENLG. In other words, these simulations show that isolated galaxies in NLG and MOG lead to faster bars compared to the model in standard dark matter paradigm. Of course, the ratio $\mathcal{R}$ obtained in our simulations for ENLG (and MOG) are still higher than the relevant observations. Nonetheless, they are much smaller than similar simulations with dark matter halo. Therefore one may conclude that we see a better agreement in the context of modified gravity theories. However, it should be emphasized that the model studied in this paper is an ideal model in the sense that we have not included the bulge component. Consequently, still more investigations with more realistic galaxy models are required to shed light on this important issue.

As our final remark, we emphasize that the dynamics of the pattern speed is a rare situation in which modified gravity and particles dark matter behave completely opposite to each other. We know that modified gravity theories are presented to play the same role as the dark matter particles to explain the rotation curves of spiral galaxies or the cosmological structure formation. But this is not the case when studying the bar evolution, and it seems that some features of particles, like the dynamical friction, cannot be recovered by modifications in the gravitational theory.
\section{\small{Discussion and Conclusion}}
\label{disc}
In this paper, we studied the evolution of a numerical galactic model in the context of a nonlocal gravity theory. More specifically, we have modified and tested the \textit{GALAXY} code to include the linear effects of NLG. Our numeric model, called ENLG, consists of an exponential disk, and its initial properties are identical to the model EPL. The model EPL consists of a spherical Plummer halo and an exponential disk. Therefore we have constructed two identical models one in NLG without dark matter halo and the other in standard paradigm including a spherical halo. The main purpose of this study is to compare the evolution and dynamics of the stellar bar in the above mentioned theories. Since the gravitational force is somehow similar to that of MOG, for the sake of completeness, we have also compared the results with MOG.

Although models start with the same initial conditions, the dynamics of the bar is substantially different. For example, we found that, at least for the exponential disks, the bar instability is stronger in NLG. More specifically, the stellar bar in nonlocal grows with a higher rate compared with the dark matter model and MOG. We recall that a rigid halo can suppress the bar instability and a live halo, in principle, can trigger and even enhance the instability \citep{at2002,se16}. From this perspective, the NLG effects appear like a live halo.

When the bar instability occurs in ENLG, and accordingly the bar reaches its maximum strength, the resonance of the particles vertical to the plane of the disk, increases the thickness. In other words, buckling instability occurs and finally, a peanut shape appears in the simulation. The peanut is meaningfully stronger than in MOG and is comparable to that of in dark matter halo model.

We also studied the long-term evolution of the stellar bar in the model ENLG. Unlike in the dark matter model, the pattern speed does not decrease with time. More specifically the dynamical friction is not effective in the ENLG disk, while it is a key feature in the dark matter halo model EPL. Finally, we measured the ratio of corotation radius to the bar semi-major axis, i.e., $\mathcal{R}$, to compare the pattern speed with the observations. We found that after $\tau\simeq 200$, $\mathcal{R}$ starts to decrease and at the end of the simulation reaches the border $\mathcal{R}=1.4$. While in the dark matter model this parameter increases with time and reaches much higher values. In other words, NLG predicts faster bars in contrary to the standard case which leads to slower bars. This totally opposite prediction along with the relevant observations may help to distinguish between modified gravity and dark matter particles.

As our final remark in this paper, for future works, it should be noted that more realistic galactic models including initial spherical bulge are required to obtain a reliable comparison between N-body simulations in modified gravity and observations. Furthermore, more accurate analysis like 2D line-of-sight kinematics of the simulations would be helpful to derive the main properties of bars and peanuts in a way suitable for comparison with observations, for example see \cite{iann,mola}. To the best of our knowledge, such analysis has not been done in modified theories of gravity. On the other hand, the cosmological simulations have not been developed in the context of modified theories of gravity without cold dark matter. Therefore a huge numerical and theoretical effort is still required to decide about the viability of these theories in the dynamics of isolated galaxies and in the cosmological galaxy formation scenario. 

\acknowledgments
The main part of the calculations have been done on ARGO cluster in International Center for Theoretical Physics (ICTP).  MR would like to thank ICTP for giving access to ARGO cluster. MR would like to appreciate Bahram Mashhoon for valuable discussions on the foundation of the non-local theory of gravity. Furthermore, we are grateful to Shahram Abbassi for providing us with a high performance computer. Also we would like to thank referee for his/her useful comments that improved this work. This work is supported by Ferdowsi University of Mashhad under Grant No. 2/47321 (02/05/1397).

\bibliographystyle{apj}
\bibliography{ENLGrev1}

\begin{thebibliography}{25}
\expandafter\ifx\csname natexlab\endcsname\relax\def\natexlab#1{#1}\fi
\bibitem[{{Aguerri} et al (2015)}]{pattern} Aguerri J. A. L. et al., 2015, A\& A, 576, A102
\bibitem[{{Algorry} et al (2017)}]{cosmo} Algorry D. G. et al., 2017, \mnras, 469, 1054
\bibitem[{{Athanassoula} (2002)}]{at2002} {Athanassoula}, E. 2002, \apj, 569, L83 
\bibitem[{{Blagojevi{\'c}} \& {Hehl} (2013)}]{tpar} Blagojevi{\'c}, M. \& Hehl, F. W., Gauge Theories of Gravitation (Imperial College Press, London, 2013).
\bibitem[{{Bohr} \& {Rosenfeld} (1950)}]{bohr} Bohr, N. \& Rosenfeld, L. 1950, Phys. Rev. 78, 794
\bibitem[{{Boylan-Kolchin}  et al (2008)}]{merge} Boylan-Kolchin, M., Ma C. \& Quataert, E. 2008, \mnras, 383, 93
\bibitem[{{Brada} \& {Milgrom} (1999)}]{brada} {Brada}, R. \& {Milgrom}, M. 1999, \apj, 519, 590
\bibitem[{{Christodoulou} (1991)}]{chris} Christodoulou, D. M. 1991, \apj, 372, 471
\bibitem[{{Combes} (2014)}]{com1} Combes F., 2014, A\&A, 571, A82
\bibitem[{{Combes} \& {Sanders} (1981)}]{cs} Combes, F. \& Sanders, R. H. 1981, A\& A, 96, 164
\bibitem[{{Debattista} \& {Sellwood} (2000)}]{deb2000} Debattista, V. P. \& Sellwood, J. A. 2000, \apj, 543, 704
\bibitem[{{Dejonghe} (1987)}]{de} Dejonghe, H. 1987, \mnras, 224, 13
\bibitem[{{Famaey} \& {McGaugh} (2012)}]{fa} Famaey, B. \& McGaugh, S.S. 2012, LRR, 15, 10
\bibitem[{{Ghafourian} \& {Roshan} (2017)}]{gr2017} Ghafourian, N. \& Roshan, M. 2017, \mnras, 468, no.4, 4450
\bibitem[{{Hohl} (1971)}]{ho} Hohl, F. 1971, \apj, 168, 343.
\bibitem[{{Iannuzzi} \& {Athanassoula} (2015)}]{iann}  Iannuzzi F., Athanassoula E., 2015, \mnras, 450, 2514
\bibitem[Jackson(1975)]{jackson} Jackson, J.~D.\ 1975, Classical Electrodynamics, New York: Wiley, 1975, 2nd ed.,  
\bibitem[{{Julian} \& {Toomre} (1966)}]{julian}
Julian, W. H., \& Toomre, A. 1966. \apj, 146, 810
\bibitem[{{Molaeinezhad} et al (2016)}]{mola} Molaeinezhad, A., et al, 2016, \mnras, 456, 692
\bibitem[{{Mashhoon} (2013)}]{mashhoon} Mashhoon, B. 2013, Class. Quantum Gravity, 30, 155008
\bibitem[{{Mashhoon} (2017)}]{mashhoonbook} Mashhoon, B. 2017, Nonlocal Gravity, Vol. 167(Oxford: Oxford Univ. Press)
\bibitem[{{Masset} \& {Tagger} (1997)}]{masset} Masset, F., \& Tagger, M. 1997, A\&A, 322, 442
\bibitem[{{Milgrom} (1983)}]{milgrom} Milgrom, M. 1983, \apj, 270, 384
\bibitem[{{Milgrom} (1989)}]{mil89} Milgrom, M. 1989, \apj, 338, 121
\bibitem[{{Miller} et al (1970)}]{miller} Miller, R. H., Prendergast, K. H., Quirk, W. J. 1970, \apj, 161, 903
\bibitem[Moffat (2006)]{m2006} Moffat, J. W. 2006, JCAP, 0603, 004
\bibitem[Moffat \& Rahvar(2013)]{2013MNRAS.436.1439M} Moffat, J.~W., \& Rahvar, S.\ 2013, \mnras, 436, 1439 
\bibitem[Moffat \& Rahvar(2014)]{2014MNRAS.441.3724M} Moffat, J.~W., \& Rahvar, S.\ 2014, \mnras, 441, 3724 
\bibitem[Monaghan (1992)]{mon1992}  Monaghan, J. J. 1992, ARAA, 30, 543
\bibitem[Ostriker \& Peebles (1973)]{op} Ostriker, J. P. \& Peebles, P. J. E. 1973, \apj, 186, 467
\bibitem[Rahvar \& Mashhoon(2014)]{rahvar} Rahvar, S., \& Mashhoon, B.\ 2014, \prd, 89, 104011 
\bibitem[\protect\citeauthoryear{Romeo}{Romeo}{1992}]{romeo1992stability} Romeo, A.~B. 1992, \mnras, 256, 307 
\bibitem[Roshan \& Abbassi (2015)]{ro2015} Roshan, M. \& Abbassi, S. 2015, \apj, 802, 9 
\bibitem[Roshan (2018)]{roshan2018} Roshan, M. 2018, \apj, 854, no. 1, 38
\bibitem[{{Sellwood} (2003)}]{se3} Sellwood, J. A. 2003, \apj, 587, 638
\bibitem[{{Sellwood} (2014)}]{se2014} Sellwood, J. A. 2014, arXiv:1406.660
\bibitem[{{Sellwood} (2016)}]{se16} Sellwood, J. A. 2016, \apj, 819, 92
\bibitem[{{Sellwood} \& {McGaugh} (2005)}]{sm2005} Sellwood, J. A. \& McGaugh, S. S. 2005, \apj, 634, 70
\bibitem[{{Tiret} \& {Combes} (2007)}]{ti} Tiret, O., \& Combes, F. 2007, \aap, 464, 517
\bibitem[{{Toomre} (1964)}]{toom64} Toomre, A. 1964, \apj, 139, 1217
\bibitem[{{Young} (1980)}]{y1980} Young, P. 1980, \apj, 242, 1232
\end{thebibliography}
\label{lastpage}
\end{document}